\documentclass{jfm}
\usepackage{graphicx}
\usepackage{newtxtext}
\usepackage{newtxmath}
\usepackage{natbib}
\usepackage{hyperref}
\usepackage{enumerate}
\usepackage{bm}
\usepackage{caption}
\usepackage{subfigure} 
\hypersetup{
    colorlinks = true,
    urlcolor   = blue,
    citecolor  = blue,
}

\newcommand{\RomanNumeralCaps}[1]

\usepackage{amsmath,amssymb}
\hypersetup{
	bookmarksnumbered,
	linkcolor=blue,
	anchorcolor = blue,
}

\graphicspath{{./Fig/},{./}}
\usepackage{placeins} 

\usepackage{cleveref}
\crefname{figure}{figure}{figures}
\crefname{equation}{}{}
\crefformat{section}{\S#2#1#3}
\crefmultiformat{section}{\S\S~#2#1#3}{ and~#2#1#3}{, #2#1#3}{, and~#2#1#3}

\usepackage{siunitx}
\usepackage{textcase}
\newcommand{\RomanNum}[1]{\MakeUppercase{\romannumeral#1}}

\title{
Streamwise dispersion with interphase mass transfer through a tube
}

\author{Mingyang Guan\aff{1}
  and Zhan Wang\aff{1,2}
 \corresp{\email{zwang@imech.ac.cn}}}

\affiliation{\aff{1}Key Laboratory for Mechanics in Fluid Solid Coupling Systems, Institute of Mechanics, Chinese Academy of Sciences, Beijing 100190, China
\aff{2} School of Engineering Science, University of Chinese Academy of Sciences, Beijing 100049, China}

\begin{document}
\maketitle

\begin{abstract}
Reactive solute transport in shear flows is of broad importance in chemical engineering, liquid--solid chromatography, and environmental mechanics, particularly in flow reactors and fluidic devices involving interphase mass transfer.
Based on the recently developed streamwise dispersion theory for inert solutions (Guan \& Chen, 2024, \textit{J. Fluid Mech.}, vol. 980, A33), we present an analytical model for reactive dispersion of chemically active solutes characterized by first-order, irreversible surface reactions.
The extension is not a simple perturbation because the reactive boundary destroys mass conservation with streamwise and radial derivatives coupled.
The model establishes how the microscopic kinetic reactivity and radial eigenmodes combine to determine a newly defined (effective) local exchange rate.
We derive rigorous explicit solutions for the local moments, concentration, and phenomenological transport coefficients at arbitrary reactivity, in quantitative agreement with numerical simulations across all regimes.
For tube Poiseuille flows, we reveal pre-asymptotic behaviours including sign-changing radial exchange, reactivity-enhanced drift, reactivity-reduced dispersivity, and non-Gaussian concentration profiles. 
Small- and large-Damköhler-number asymptotics for the local and global transport coefficients identify the reaction-limited and diffusion-limited regimes, connecting the conservative case to the Dirichlet limit at infinite reactivity.
This streamwise perspective could advance our understanding of macroscale transport in chemically active systems.
\end{abstract}


\section{Introduction}
\label{sec:intro}
Reactive solute transport in tubes arises in flow reactors \citep{balakotaiahDispersionChemicalSolutes1995,chazot-franguiadakisOpticalQuantificationNanopores2022}, chromatography \citep{haweTaylorDispersionAnalysis2011,aquinoEquilibriumDistributionsAdvection2024}, and environmental transport \citep{zhangAnalyticalSolutionSolute2025}.
In confined heterogeneous systems, convection and molecular diffusion are coupled to interphase mass transfer at the wall, so the transient concentration field depends on both the velocity profile and the history of radial exchange \citep{lingDispersionControlledPermeable2016,lingDispersionControlCoupled2024}. 
The principal difficulty is that wall reaction destroys mass conservation. The streamwise formulation developed by \cite{guanStreamwiseDispersionSoluble2024} for inert solutes therefore cannot be extended by perturbation.
Although cross-sectionally averaged or global-moment descriptions can capture overall mass exchange, drift and dispersivity \citep{gillExactAnalysisUnsteady1970,deganceTheoryDispersionChemically1978,bartonAsymptoticTheoryDispersion1984,jiangAnalyticalSolutionsReactive2022}, they do not retain the streamwise dispersion history that governs the local concentration field.

For inert solutes, the classical short- and long-time limits are well established.
At an early time $t$ after a point-source release in a free space, the mean concentration distribution $\bar{C}$ of passive particles over the cross-section along the flow direction $x$ is locally diffusive \citep{saffmanEffectMolecularDiffusivity1960,latiniTransientAnomalousDiffusion2001}:
\begin{eqnarray} 
	\lim\limits_{t\to 0}\bar{C} (x,t) \simeq  \frac{1}{\sqrt{2\upi \sigma^2}} \exp \left[
	-\frac{(x-\Pen u(r_0) t)^2}{2\bar\sigma^2}
	\right],\label{eq_short_meanC}
\end{eqnarray}
with the mean variance $\bar\sigma^2 \sim 2t$.
Here, $u(r_0)$ is the flow velocity of the release position, and $\Pen$ is the dimensionless P\'{e}clet number, defined as the ratio of convection to diffusion.
For sufficiently long times, radial diffusion homogenizes the cross-section and the concentration approaches the Taylor-dispersion limit \citep{taylorDispersionSolubleMatter1953}:
\begin{eqnarray}
	\lim\limits_{t\to \infty} \bar{C} (x,t) \simeq \frac{1}{\sqrt{2\upi \bar\sigma^2}} \exp \left[
	-\frac{(x-\Pen \bar{u} t)^2}{2\bar\sigma^2}
	\right],\label{eq_long_meanC}
\end{eqnarray}
with $\bar\sigma^2 \sim 2 \bar{D}_\text{eff} + \text{const}$.
Here, $\bar{u}$ is the mean flow speed and $\bar{D}_\text{eff}$ is the effective diffusivity (Taylor dispersivity).

The unsteady dispersion theory of Gill and Sankarasubramanian \citep{gillNoteSolutionTransient1967,gillExactAnalysisUnsteady1970} provides a natural description between these limits:
\begin{eqnarray}
	\bar{C} (x,t) \simeq \frac{1}{\sqrt{2\upi [\bar\sigma(t)]^2}} \exp \left[
	-\frac{(x- \bar{\mu}(t) t)^2}{2[\bar\sigma(t)]^2}
	\right],\label{smith_1term_Gaussian}
\end{eqnarray}
where $\bar\mu(t)$ is the time-dependent mean centroid.
However, a cross-sectionally averaged representation does not by itself resolve the distinct transport histories carried by different streamlines \citep{smithGAUSSIANAPPROXIMATIONCONTAMINANT1982}.
During the transitional regime, local centroids, variances and higher moments may evolve differently, leading to non-Gaussian concentration fields, including skewed and transient double-peaked profiles \citep{sullivanLongitudinalDispersionTwodimensional1971,yasudaLongitudinalDispersionMatter1984}. These effects motivate retaining the local streamwise statistics rather than imposing cross-sectional homogenization at the outset.

This perspective motivates a streamwise, or Lagrangian, framework of \cite{guanStreamwiseDispersionSoluble2024}.
For an inert solute in tube Poiseuille flows, a coordinate convected decouples convection from the bulk transport equation and yields an exact diffusion-type problem with a time-dependent anisotropic diffusivity tensor.
The formulation preserves the streamwise transport history while recovering Taylor's long-time solution \citep{taylorDispersionSolubleMatter1953}.
Its short-time solution takes the form,
\begin{eqnarray}
	C(x,r,t) = \frac{2\upi}{\left(4\upi t\right)^{3/2}} \exp \left[
	-\frac{(x- u(r) t)^2+r^2}{4t}
	\right],
	\label{sol_guan_particular}
\end{eqnarray}
which provides a local benchmark for the subsequent evolution, complementary to the classcal Taylor--Aris result.

The reactive problem introduces a qualitatively different mathematical structure \citep{brennerMacrotransportProcesses1993,shapiroChemicallyReactiveGeneralized1987,biswasTaylorDispersionAbsorbing2007,jiangSolutionGillGeneralized2018}. 
In the same convected coordinate, convection is removed from the bulk equation, but the first-order reactive Robin condition couples the streamwise and radial derivatives. 
The derivative coupling is induced directly by the boundary mass flux and has no counterpart in the conservative problem. 
Thus, the non-conservative boundary condition remains coupled to the streamwise evolution even after the bulk convection has been transformed away. 
This coupling is the central mathematical difficulty in constructing a local streamwise theory for reactive dispersion.

Recently, nanoscale Taylor dispersion has been shown to exhibit a striking reduction in spreading, with the dynamics slowed by a factor of ten relative to the conservative case \citep{vilquinNanoparticleTaylorDispersion2023}.
The corresponding physical change is captured by the radial eigenstructure. 
For an inert solute, the lowest radial eigenvalue is zero, reflecting mass conservation; for an irreversibly reacting solute, it becomes positive, producing global mass loss while radial diffusion continues to redistribute solute across the tube.
Solute in the core can be locally depleted, whereas the near-wall region can exhibit positive local exchange because solute is redistributed radially towards the reactive boundary.
This sign change does not represent negative molecular diffusivity or negative effective diffusivity; it is a transient consequence of redistribution within a non-conservative system.

Reactive Taylor dispersion has long been studied using moment methods, eigenfunction expansions, and asymptotic analyses \citep{gillExactAnalysisUnsteady1970,sankarasubramanianUnsteadyConvectiveDiffusion1973,shapiroTaylorDispersionChemically1986,ngDispersionOscillatoryCouette2005,aquinoEquilibriumDistributionsAdvection2024}.
In particular, Barton's asymptotic analyses \citep{bartonAsymptoticTheoryDispersion1984} identify limiting behaviours for reactive dispersion, while higher-order moment descriptions have been developed for transitional regimes \citep{zhangTransientSoluteTransport2017,wangBasicCharacteristicsTaylor2017,jiangAnalyticalSolutionsReactive2022}. 
These approaches establish important global transport characteristics, but do not provide the local, time-resolved description required to resolve the redistribution mechanism described above. 
A local formulation is also relevant to the interpretation of Taylor-dispersion measurements, for which wall reaction can alter the inferred transport coefficients \citep{haweTaylorDispersionAnalysis2011}.

Here, we extend the streamwise framework of \cite{guanStreamwiseDispersionSoluble2024} to irreversible first-order reaction. 
We derive explicit solutions for the local concentration and local moments, and from them obtain phenomenological transport coefficients at arbitrary reactivity. 
The resulting theory resolves the pre-asymptotic regimes controlled jointly by the Péclet and Damköhler numbers, and shows that increasing reactivity enhances the effective drift while reducing dispersivity. 
It also identifies the sign-changing radial exchange rate and explains its relation to transient radial redistribution. 
The predicted concentration fields reproduce the non-Gaussian transitional behaviour found in numerical simulations, including skewed and double-peaked profiles, and show quantitative agreement with the numerical results. 
In addition, small- and large-Damköhler-number asymptotic analyses identify reaction-limited and diffusion-limited regimes, and show how the reactivity continuously connect the conservative limit to the Dirichlet limit of infinitely fast wall reaction.

The remainder is organized as follows.
In §\ref{formula}, we revisit the non-conservative dispersion problem and derive the local and global moment solutions and the associated transport coefficients in terms of the radial eigenmodes.
Section \ref{model} develops the streamwise dispersion model for chemically active solutes.
In §\ref{phys}, we examine the underlying mechanisms through flux vectors, and identify the relevant dispersion regimes through scale analyses.
Section \ref{result} presents the local and global transport coefficients, their asymptotics, and comparisons with numerical simulations for representative source distributions, with emphasis on the transient concentration profiles and the limiting regimes. 
Finally, \S\ref{conclusion} summarizes the main findings and the connections between conservative, reaction-limited, diffusion-limited, and Dirichlet limits.
Technical details of the moment solutions and numerical simulations are given in appendices \S\ref{appA} and \S\ref{appB}.

\section{Formulation}
\label{formula}
\subsection{Non-conservative transport of reactive solute}
\begin{figure}
	\centerline{\includegraphics{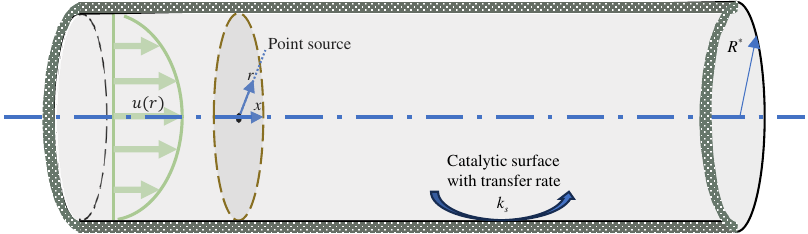}}
	\caption{Streamwise dispersion with interphase mass transfer for a central point-release in a tube Poiseuille flow. The chemically active solutes reacts at the catalytic wall through an irreversible first-order surface reaction characterized by $k_s$.}
	\label{fig_sketch}
\end{figure}

We consider the dispersion of a reactive solute in a straight, circular tube, where the chemical reaction occurs at the tube wall, as illustrated in \cref{fig_sketch}. 
The wall is impermeable to the fluid, and removes solute through an irreversible first-order surface reaction.
The flow is a fully developed, steady, laminar Poiseuille flow, such that the axial velocity $u$ depends only on the radial coordinate,
\begin{equation}
	u(r) = 2\bigl(1-r^{2}\bigr),
\end{equation}
The solute concentration is sufficiently dilute that the fluxes of different chemical species are uncoupled, and the transport is axisymmetric.

In dimensionless form, the azimuthally averaged concentration field $C=C(x,r,t)$ satisfies the convection--diffusion equation as
\begin{eqnarray}
	\frac{\partial C}{\partial t} + \Pen ~ u(r) \frac{\partial C}{\partial x} = \frac{\partial^2 C}{\partial x^2} + \frac{1}{r} \frac{\partial }{\partial r} \left(r\frac{\partial C}{\partial r}\right).
	\label{eq_cde}
\end{eqnarray}
The solute is released instantaneously at $x=0$ with a prescribed radial distribution $Y(r)=\delta(r-r_0)/2r_0$ corresponding to a ring source located at $r=r_0\neq0$, or a central point source for $r_0=0$.
The initial condition is therefore
\begin{eqnarray}
	C(x,r,t=0) = \delta(x) Y(r),
\end{eqnarray}
where $\delta$ denotes the Dirac delta function.
The normalization of $Y$ is such that the initial dimensionless mass is unity.
Since the governing equation is linear, the localized pulse provides a Green-function solution from which arbitrary initial or inlet conditions can be constructed by superposition.

At the tube wall, $r=1$, the diffusive flux toward the wall is balanced by the rate of interphase mass transfer, which is characterized by an irreversible first‑order reaction. 
Combining Fick's law with linear surface kinetics gives the Robin boundary condition
\begin{eqnarray}
	\frac{\partial C}{\partial r} = - k_s C(x,r,t),\quad  \text{on the tube wall } r=1,
	\label{bc_wall}
\end{eqnarray}
where the Damköhler number $k_s$ measures the ratio of the characteristic reaction velocity to the diffusive velocity.

The concentration remains finite on the centreline,
\begin{equation}
	\left.C(x,r,t)\right|_{r=0} \quad \text{finite},
\end{equation}
and the solute cloud decays sufficiently rapidly in both axial directions such that
\begin{equation}
	\left.
	x^n\frac{\partial^n C}{\partial x^n}
	\right|_{x\to\pm\infty}
	=0,
	\qquad n=0,1,2,\ldots.
\end{equation}

The dimensionless variables are related to their dimensional counterparts, denoted by an asterisk, according to
\begin{equation}
	C = \frac{C^\ast \upi R^{\ast 3}}{m^\ast},\quad
	t = \frac{t^\ast D^\ast}{R^{\ast 2}},\quad
	\Pen = \frac{\bar{u}^\ast R^\ast}{D^\ast},\quad
	u = \frac{u^\ast}{\bar{u}^\ast},\quad
	x = \frac{x^\ast}{R^\ast},\quad
	r = \frac{r^\ast}{R^\ast},\quad
	k_s = \frac{k_s^\ast R^\ast}{D^\ast},
\end{equation}
where $C^\ast$ is the local concentration, $R^\ast$ is the tube radius, $m^\ast$ is the initial total mass of solute released, $t^\ast$ is the time, $D^\ast$ is the Brownian diffusivity, $u^\ast$ and $\bar{u}^\ast$ are the local and mean flow velocities, respectively, $x^\ast$ and $r^\ast$ are the axial and radial coordinates, and $k_s^\ast$ is the kinetic surface reactivity.
The concentration scale is chosen such that the volume integral of the initial release gives the total mass exactly.

\subsection{Local moments of concentration}\label{sec:local_moment}

In the spirit of the Taylor--Aris dispersion framework, we characterize the spatialtemporal evolution using axial moments of the concentration. 
The local (radially dependent) axial moments of order $n$ and the corresponding global moments are defined, respectively, as 
\begin{eqnarray}
	&C_n(r,t) = \int_{-\infty}^{\infty} x^n C ~ \mathrm{d}x,\\[8pt]
	&M_n(t) = \int_0^1 C_n ~ 2r~\mathrm{d}r = \int_{-\infty}^{\infty} x^n \bar{C} ~ \mathrm{d}x,\quad n=0,1,2,\ldots,
	\label{eq_def_moments}
\end{eqnarray}
where $\bar{C}=\int_0^1 C ~2r~\mathrm{d}r$ is the mean concentration.
The zeroth global moment $M_0$ represents the total dimensionless mass remaining in the system and therefore decays in time when $k_s>0$.
The first and second global moments $M_1$ and $M_2$ determine the mean centroid and variance of the solute cloud after normalization by the remaining mass, while higher-order moments describe increasingly detailed features of the axial distribution, including skewness and kurtosis.
The local moments, in contrast, retain the dependence on radial position and therefore resolve the dispersion associated with different radial locations in the cross-section.

Multiplying the convection–diffusion equation (\ref{eq_cde}) by $x^n$ and integrating over $x\in(-\infty,\infty)$, with the far-field conditions used to eliminate the axial boundary terms, gives the hierarchy:
\begin{eqnarray}
	\left. \begin{aligned}
		& \frac{\partial {{C}_{n}}}{\partial t}-n\Pen u{{C}_{n-1}}=n(n-1){{C}_{n-2}}+\frac{1}{r}\frac{\partial }{\partial r}\left( r\frac{\partial {{C}_{n}}}{\partial r} \right), \\[8pt]
		& \frac{\partial {{C}_{n}}}{\partial r}=-{{k}_{s}}{{C}_{n}},\quad \text{on the tube wall }r=1, \\[8pt]
		& {{\left. {{C}_{n}}(r,t) \right|}_{r=0}}=\text{ finite}, \\[8pt]
		& {{C}_{n}}(r,t=0)=\left\{ \begin{matrix}
			Y(r), \quad~~~~ n=0,  \\
			0,\quad n=1,2,\ldots,   \\
		\end{matrix} \right. \\ 
	\end{aligned} \right\}
	\label{eq_moment_hierarchy}
\end{eqnarray}
with $C_{-1}=C_{-2}=0$.
The homogeneous Robin condition is inherited directly from (\ref{bc_wall}); since it is non-conservative for $k_s>0$, it leads to the decay of $M_0$.

The radial differential operator, together with the regularity condition at $r=0$ and the Robin boundary condition at $r=1$, forms a self-adjoint Sturm--Liouville problem \citep{bartonAsymptoticTheoryDispersion1984,jiangAnalyticalSolutionsReactive2022}.
The moment hierarchy can therefore be solved by expansion in the corresponding orthonormal eigenfunctions. The eigenvalue problem is
\begin{eqnarray}
	\left. \begin{aligned}
	&\frac{1}{r} \frac{\partial}{\partial r} \left(r\frac{\partial f}{\partial r}\right)
	= -\lambda f, \\[8pt]
	& \frac{\partial f}{\partial r} = -k_s f,\\[8pt]
	& \left. f \right|_{r=0} = \text{finite},
\end{aligned} \right\}
\end{eqnarray}
where $\lambda$ is the eigenvalue and $f(r)$ is the corresponding eigenfunction.

For $k_s\geq 0$, the admissible eigenvalues are positive, satisfying the transcendental equation:
\begin{eqnarray}
	\sqrt{\lambda} ~ {\mathrm{J}_1 (\sqrt \lambda)} = {k_s} {\mathrm{J}_0 (\sqrt \lambda)},
\end{eqnarray}
which defines the discrete spectrum $\{\lambda_i\}_0^\infty$=$\{\alpha_i^2\}_0^\infty$.
For the conservative case $k_s=0$, the lowest eigenvalue is $\lambda_0=0$, corresponding to the constant radial eigenfunction, and the problem reduces to the classical conservative case.

For $k_s>0$, the corresponding normalized eigenfunctions are
\begin{eqnarray}
	f_i = \dfrac{\sqrt{2 \lambda_i} \mathrm{J}_0 (\sqrt{\lambda_i} r)}{\mathrm{J}_0 (\sqrt{\lambda_i}) \sqrt{k_s^2+\lambda_i}},
\end{eqnarray}
where $\mathrm{J}_i$ is the Bessel function of the first kind of order $i$.
The normalization is chosen to satisfy $\overline{f_i f_j} \triangleq \int_0^1 f_i f_j r\mathrm{d}r=\delta_{i\!j}$, where this overline operator for eigenfunctions differs from the overbar used in mean concentration $\bar{C}$ by a factor of 2.

The analytical treatment of moments beyond second order in the presence of surface reaction was historically cumbersome, and only relatively recently have systematic closed-form results become available \citep{biswasTaylorDispersionAbsorbing2007}.
In particular, analytical expressions for the first few moments, including mass decay, centroid displacement and variance, have been derived \citep{wangBasicCharacteristicsTaylor2017,zhangTransientSoluteTransport2017,jiangAnalyticalSolutionsReactive2022}.
Explicit solutions for the first few orders are given in Appendix \ref{appA}, where the local moment equations are solved first and subsequently averaged over the cross-section.

Following the approach of \cite{bartonMethodMomentsSolute1983,bartonAsymptoticTheoryDispersion1984}, we obtain a useful solution for the zeroth axial moment, which corresponds to the Smoluchowski-type radial diffusion--reaction problem associated with (\ref{eq_cde}).
The solution is
\begin{equation}
	C_0 (r,t) = 
	\sum\limits_{i=0}^{\infty} \frac{2 \lambda_i \int_0^1 Y(r) \mathrm{J}_0 (\sqrt{\lambda_i} r) r\mathrm{d}r}{(k_s^2+\lambda_i)\mathrm{J}_0^2(\sqrt{\lambda_i})} 
	\mathrm{J}_0(\sqrt{\lambda_i} r)
	\exp({-\lambda_i t}),
\end{equation}
and on substituting the central point source condition $Y(r)=\delta(r-r_0)/(2r_0)$ as the limit $r_0\to0$,
\begin{equation}
	C_0(r,t)
		=
	\sum\limits_{i=0}^{\infty} 
	\dfrac{\lambda_i \mathrm{J}_0(\sqrt{\lambda_i} r)}{(k_s^2+\lambda_i)\mathrm{J}_0^2(\sqrt{\lambda_i})}
	\exp({-\lambda_i t}).
\end{equation}
This solution also serves as a Green function for arbitrary initial conditions.
The zeroth global solution, obtained by cross-sectional integration, is
\begin{equation}
	M_0(t) = 2 k_s \sum\limits_{i=0}^{\infty} 
	\dfrac{\mathrm{J}_{0} (\sqrt{\lambda_i r_0}) \exp({-\lambda_i t})}{(k_s^2+\lambda_i)\mathrm{J}_0(\sqrt{\lambda_i})}
	=2 k_s \sum\limits_{i=0}^{\infty} 
	\dfrac{\exp({-\lambda_i t})}{(k_s^2+\lambda_i)\mathrm{J}_0(\sqrt{\lambda_i})},
	\label{sol_M0_ring_text}
\end{equation}
for a central point source.
For a uniform area source, the corresponding solution is
\begin{equation}
	M_0(t) 
	=4 k_s^2 \sum\limits_{i=0}^{\infty} 
	\dfrac{\exp({-\lambda_i t})}{\lambda_i(k_s^2+\lambda_i)}.
	\label{sol_M0_area_text}
\end{equation}
These series converge rapidly for $t>0$. In the limit $k_s\to0$ corresponding to an impermeable, non-reactive wall, the zeroth global moment tends to unity, consistent with mass conservation. For $k_s>0$, $M_0\to0$ as $t\to\infty$.
The above explicit solutions provide useful benchmarks for numerical validation.

Finally, expanding the local moments in the same eigenfunction basis,
\begin{eqnarray}
	C_n = \sum\limits_{i=0}^\infty p_{ni}(t) \exp (-\lambda_i t) f_i(r),
\end{eqnarray}
reduces the coupled partial differential equations (\ref{eq_moment_hierarchy}) to a set of ordinary differential equations for the time‑dependent coefficients $p_{ni}$.

\section{Streamwise dispersion model for chemically active solute}
\label{model}
The extension of the streamwise framework to a chemically active solute is non-trivial because the reactive wall breaks mass conservation and, after transformation to convected coordinates, couples the streamwise and radial derivatives at the boundary.
The bulk equation can still be transformed to a diffusion equation with a time-dependent anisotropic diffusivity tensor, but the boundary condition now carries the effect of the irreversible surface reaction.
This distinction is central to the formulation below.

To separate convection from diffusion, we introduce the convected coordinate $x^\prime=x-\Pen u(r)t$, with $r^\prime=r$. The governing equation can then be written as a purely diffusive equation with an anisotropic diffusivity tensor,
\begin{eqnarray}
	\frac{\partial C}{\partial t} =  \bnabla^\prime \bcdot  \left(\mathsfbi{D} ~ \bnabla^\prime C\right),
	\quad 	\mathsfbi{D}=\mathsfbi{{D}^{(O)}}+{\Pen~{ t}}\mathsfbi{{D}^{({C})}}+{{\left( {\Pen~{ t}} \right)}^{2}}\mathsfbi{{D}^{(E)}},\label{con_Tao_par}
\end{eqnarray}
where
\begin{equation}
	\mathsfbi{{D}^{({O})}}=\left[ \begin{array}{*{35}{l}}
		1 & 0   \\[8pt]
		0 & 1   
	\end{array} \right],\quad \mathsfbi{{D}^{({C})}}=\left[ \begin{matrix}
		0 & -\frac{\partial u}{\partial r^\prime}  \\[8pt]
		-\frac{\partial u}{\partial r^\prime} & 0
	\end{matrix} \right],\quad \mathsfbi{{D}^{({E})}}=\left[ \begin{matrix}
		\left( \frac{\partial u}{\partial r^\prime} \right)^{2} & 0  \\[8pt]
		0 & 0   
	\end{matrix} \right].
	\label{coe_diffu}
\end{equation}
Thus, $\mathsfbi{D}^{(O)}$, $\mathsfbi{D}^{(C)}$ and $\mathsfbi{D}^{(E)}$ represent, respectively, ordinary molecular diffusion, shear-induced cross-diffusion and the quadratic shear enhancement responsible for Taylor dispersion. 
Here, for the axisymmetric $(x',r')$ coordinates, $\bnabla^\prime= (\partial_x^\prime, \frac{1}{r^\prime}\partial_r^\prime)$, with the cylindrical form of the divergence understood.

The essential modification is the catalytic wall, which imposes the reactive Robin condition (\ref{bc_wall}) and renders the transport non-conservative.
Applying the same transformation in (\ref{bc_wall}) gives
\begin{equation}
	\frac{\partial C}{\partial r^\prime}
	-\Pen t \frac{\partial u}{\partial r^\prime}
	\frac{\partial C}{\partial x^\prime}
	=
	-k_s C,
	\qquad \text{at }r^\prime=1.
	\label{convected_robin_bc}
\end{equation}
Thus, the transformation that removes convection from the bulk equation does not decouple the reactive boundary: the Robin condition retains an explicit coupling between the streamwise and radial derivatives.

The non-conservative character also changes the structure of the radial eigenproblem.
For $k_s=0$, the lowest eigenvalue is $\lambda_0=0$, reflecting conservation of total mass, whereas for $k_s>0$ the lowest eigenvalue satisfies $\lambda_0(k_s)>0$.
The mode associated with $\lambda_0$ therefore controls the long-time global mass decay, while the first non-zero radial mode, characterized by $\lambda_1$, controls the relaxation of the radial concentration profile towards its asymptotic form.
Reaction thus introduces a second role for the radial spectrum beyond cross-sectional homogenization: the slowest mode becomes directly responsible for irreversible mass loss.

The resulting evolution can be described using the local streamwise concentration distribution without repeating the derivation of \cite{guanStreamwiseDispersionSoluble2024}. 
At each radial position, we define the normalized marginal density
\begin{eqnarray}\label{con_pdf_def}
	P(x,~r,~t|~x_0,~r_0 )=\frac{C(x,~r,~t)}{C_0(r,~t)},
\end{eqnarray}
together with its raw and central moments,
\begin{eqnarray}
	\gamma_n = \int_{-\infty}^{\infty} x^n P \mathrm{d} x = \frac{C_n}{C_0},
	\quad \mu_n = \int_{-\infty}^{\infty} (x-\gamma_1)^n P \mathrm{d} x.
\end{eqnarray}
In particular, $\gamma_1$ is the local centroid, while $\mu_2$ is the local variance.
Because of the wall reaction, $C_0(r,t)$ is not conserved and generally decays in time.
Consequently, $P$ is a conditional probability density: it describes the streamwise distribution of the solute that remains at a given radial position, rather than the absolute local mass itself.
The normalization by $C_0$ is therefore essential for separating local streamwise spreading from local gain or loss caused by reaction and radial redistribution.

The Fourier transform of $P$ with respect to $x$ defines the characteristic function:
\begin{eqnarray}\label{con_charac_def}
	\phi(k,~r,~t)
	= \int_{-\infty}^{\infty} P \exp (\mathrm{i} k x)~\mathrm{d}x
	=\sum\limits_{n=1}^{\infty }{\frac{\left(\mathrm{i}k\right)^n}{n!} \dfrac{C_n}{C_0}}.
\end{eqnarray}
The cumulant‑generating function is the natural logarithm of the characteristic function,
\begin{eqnarray}
	\ln \phi =\sum\limits_{n=1}^{\infty }{\kappa_n}\dfrac{\left(\mathrm{i}k\right)^n}{n!},
	\label{con_chagen_def}
\end{eqnarray}
where $\kappa_n$ are the local cumulants with
$\kappa_1=\gamma_1$,
$\kappa_2=\mu_2$,
$\kappa_3=\mu_3$, and
$\kappa_4=\mu_4-3\mu_2^2$.

Introducing the locally centred and normalised coordinate,
$X=({{ x-{{\kappa}_{1}} }})/\sqrt{\kappa_2}$,
the inverse Fourier representation yields the streamwise dispersion model
\begin{eqnarray}
	C=\frac{{C}_{0}\exp(-X^2/2)}{\sqrt{2\pi {{\kappa }_{2}}}}
	\left[1+ 
	\frac{\kappa_3}{3! \left(\kappa_2\right)^{3/2}} \mathit{He}_{3}(X)
	+ \frac{\kappa_4}{4! \kappa_2^2} \mathit{He}_{4} (X)
	+ \frac{10\kappa_3^2}{6! \kappa_2^3} \mathit{He}_{6} (X)
	+\ldots
	\right],
	\label{con_streamwiseExpand_sol}
\end{eqnarray}
where $\mathit{He}_{n}$ is the $n$th-order Hermite polynomial. 
Physically, the cumulants provide a compact description of the local streamwise transport: $\kappa_1$ specifies the local mean position, $\kappa_2$ the leading-order spreading, and $\kappa_3$ and $\kappa_4$ the leading non-Gaussian corrections. The centring and scaling of $X$ therefore isolate the dispersive structure in a coordinate system that follows the local mean motion and is normalised by the local dispersion width.

The following section uses the corresponding flux vectors and scaling analyses to identify the mechanisms underlying the sign-changing exchange rate, the reaction-enhanced drift, the reaction-reduced dispersion, and the non-Gaussian pre-asymptotic behaviour.

\section{Reactive dispersion: mechanisms and scalings}
\label{phys}
\subsection{Flux-vector balance mechanisms}
The principal difference introduced by the catalytic wall is revealed most directly by examining the diffusive flux at the reactive boundary. 
In the convected coordinates, the flux vector is
\begin{equation}
	\boldsymbol{j}=
		\left[\begin{array}{c}
		j_x^\prime \\[8pt]
		j_r^\prime
	\end{array}\right]
	=
	\left[\begin{array}{c}
		-\left(1+16~\Pen^2~t^{2}r^{\prime 2}\right) \dfrac{\partial C}{\partial x^\prime}
		- 4~\Pen~t r^\prime~\dfrac{\partial C}{\partial r^\prime} \\[8pt]
		-\dfrac{\partial C}{\partial r^\prime}
		-4~\Pen~t r^\prime~\dfrac{\partial C}{\partial x^\prime} 
	\end{array}\right],
	\label{con_Tao_sol_flux_coo}
\end{equation}
which follows directly from $\boldsymbol{j}=-\mathsfbi{D}\cdot\bnabla^\prime C$, with the tube Poiseuille flow profile substituted.
At the reactive wall,
\begin{eqnarray}
	j_r^\prime = k_s C, \quad \text{ at } r=1,
\end{eqnarray}
so that the no-flux condition of the inert case is replaced by a finite outward flux determined by the first-order surface reaction.

For a ribbon released near the wall, the two axial ends therefore respond differently to the reactive boundary condition (\ref{convected_robin_bc}). 
At the front, $\partial_{x^\prime}C<0$, giving
\begin{eqnarray}
	\frac{\partial C}{\partial r^\prime} = \underbrace{(-4 \Pen~t~r^\prime \frac{\partial C}{\partial x^\prime})}_{>0} - \underbrace{k_s C}_{>0}, \quad \text{at the front end},
\end{eqnarray}
so that weak reaction retains the shear-driven tendency to approach the wall, whereas sufficiently strong reaction reverses the radial gradient and produces a near-wall depletion layer.
At the rear, $\partial_{x^\prime} C>0$, and hence
\begin{eqnarray}
	\frac{\partial C}{\partial r^\prime} = \underbrace{(-4 \Pen~t~r^\prime \frac{\partial C}{\partial x^\prime})}_{<0} - \underbrace{k_s C}_{>0}, \quad \text{at the rear end},
\end{eqnarray}
so that the two contributions reinforce one another and the concentration gradient is directed towards the tube axis. 
The catalytic wall therefore accentuates the asymmetry between the leading and trailing parts of the ribbon: the leading edge is preferentially depleted near the wall, whereas the rear is biased towards the interior.

For an inert solute, the corresponding boundary condition is
${\partial_{r^\prime} C}
	+4\Pen t r^\prime
	{\partial_{x^\prime} C}
	=0,$
so that the wall merely redirects the shear-induced flux. 
The reactive boundary instead permits a finite outward flux and consequently alters both the ribbon shape and the total mass. 
At long times, the mass decay is governed by the slowest reactive eigenmode. 
This additional loss mechanism modifies the transient approach to the Taylor-dispersion regime.

The effect of the wall can also be viewed through the streamwise flux in (\ref{con_Tao_sol_flux_coo}). 
The streamwise diffusivity increases with the shear contribution $\Pen^2 t^2(\partial_r u)^2$, whereas solute close to the wall is simultaneously removed by the reaction. 
Thus, shear enhancement is weak near the centreline, where $\partial_r u=0$, but is strongest near the wall, where it is also most strongly attenuated by interfacial mass transfer. 
The resulting effective dispersivity is therefore sensitive to $k_s$: its long-time value is reduced as the wall becomes more reactive, while its transient evolution along each streamline can remain non-monotonic. 

\begin{figure}
	\centerline{\includegraphics{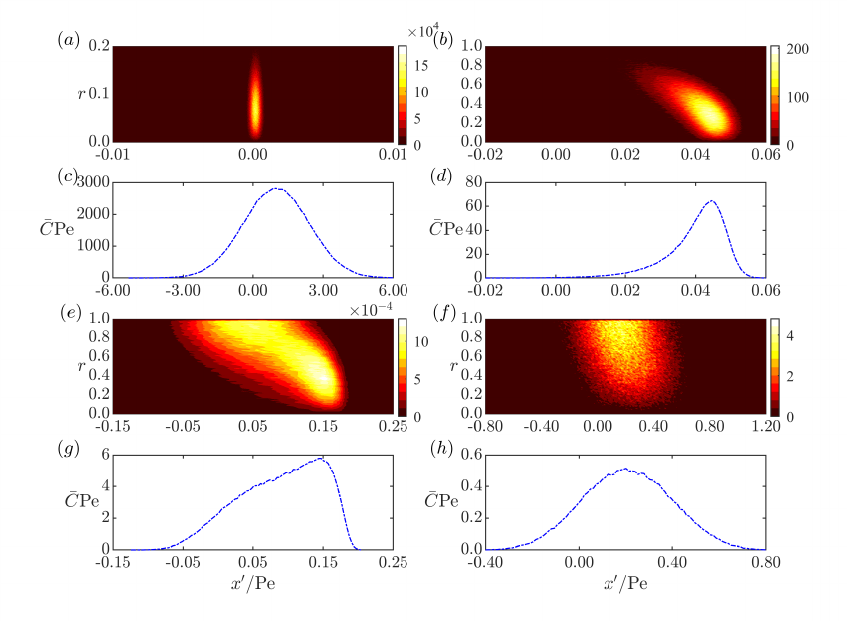}}
	\caption{Transient dispersion regimes with interphase mass transfer. The concentration distributions of a reactive solute through a catalytic tube are calculated using Monte Carlo simulations for a central point source. ($a,c$) The diffusive regime at $t=10^{-4}$. ($b,d$) The anomalous diffusive regime at $t=0.05$. ($e,g$) The pre-asymptotic regime at $t=0.2$. ($f,h$) The Taylor dispersion regime at $t=1$.
	The parameters are $k_s=1$, $\Pen=100$, $N=5\times10^{5}$, and $\Delta t=10^{-5}$.}
	\label{fig_contour}
\end{figure}

Figure \ref{fig_contour} illustrates distinctive transient regimes: diffusive, anomalous, pre-asymptotic, and Taylor dispersion. The same sequence is present for inert solutes, consistent with \cite{guanStreamwiseDispersionSoluble2024}, but the reactive wall changes the characteristic spatial and temporal scales.
The concentration field evolves from a compact, nearly symmetric distribution to a skewed structure, and ultimately to a broader, radially mixed state.
The corresponding cross-sectional concentration profile develops pronounced asymmetry in the intermediate regimes before recovering an approximately diffusive form at long times, while its amplitude decreases continuously because of mass consumption.
The reactive boundary therefore both redistributes and removes solute, accelerating the transient streamwise transport relative to the purely inert case.
The numerical results provide the physical basis for the characteristic timescales considered below.

\subsection{Regimes and scaling laws}

To quantify the temporal evolution of axial dispersion, we introduce the local logarithmic slope
\begin{eqnarray}
	\alpha(t) = \frac{\mathrm{d}\ln\bar\sigma^2}{\mathrm{d}\ln t} = \frac{t}{\bar\sigma^2} \frac{\mathrm{d}\bar\sigma^2}{\mathrm{d} t},
	\label{def_alpha}
\end{eqnarray}
which equals the exponent $\alpha$ exactly when $\bar{\sigma}^2(t) \propto t^\alpha$ and otherwise represents the instantaneous scaling exponent.
Thus, $\alpha=1$ corresponds to normal diffusion, $\alpha<1$ to subdiffusive scaling and $\alpha>1$ to superdiffusive scaling.

\begin{figure}
	\centerline{\includegraphics{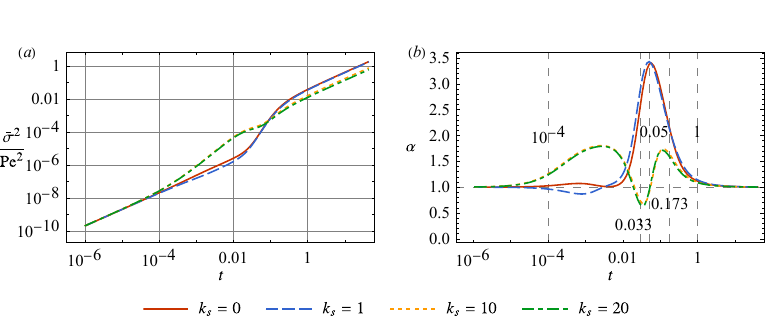}}
	\caption{Temporal evolution of ($a$) the mean axial dispersion $\bar{\sigma}^2/\Pen^2$ and ($b$) the local slope index $\alpha$ for a central point source at $\Pen=100$.}
	\label{fig_scaling}
\end{figure}

The conservative limit, $k_s=0$, recovers the classical regimes associated with the development of Taylor dispersion \citep{latiniTransientAnomalousDiffusion2001}. 
At early times, radial molecular diffusion dominates and $\bar{\sigma}^2\propto t$, giving $\alpha\simeq1$. 
At intermediate times, however, shear converts the evolving radial distribution into enhanced streamwise spreading, producing a pronounced anomalous regime with $\alpha>2$. 
This superballistic growth of the variance is consistent with the transient anomalous-diffusion picture of \cite{latiniTransientAnomalousDiffusion2001}, for which a point release in Poiseuille flow exhibits an intermediate regime between the initial diffusion-dominated stage and the long-time Taylor regime; see, e.g., figures~2(e) and 2(f) therein. 
At sufficiently long times, radial diffusion homogenizes the tracer across the cross-section and $\alpha$ approaches unity again, signalling the classical Taylor-dispersion limit. 
Thus, both the variance evolution and the local exponent in figure~\ref{fig_scaling} reproduce the essential features of the three regimes identified in earlier analyses of shear dispersion
\citep{taylorDispersionSolubleMatter1953,lighthillInitialDevelopmentDiffusion1966,youngShearDispersion1991,latiniTransientAnomalousDiffusion2001}.

For $k_s>0$, the relevant regimes are modified by irreversible reaction at the wall. 
As $k_s$ increases, the anomalous peak in $\alpha(t)$ is progressively shifted and suppressed, indicating that wall exchange increasingly constrains the transient shear-dispersion process.
The approach to complete cross-sectional mixing is consequently altered, and the subsequent decay of the transient modes leads to a $k_s$-dependent long-time effective dispersivity. 
The competition between radial diffusion, shear-induced axial spreading, and interphase mass transfer therefore provides a natural extension of the inert-boundary picture, yielding the following dispersion regimes.

\subsubsection{Regime \RomanNum{1}: Early diffusive regime ($t \ll t_c^{o} \sim \Pen^{-2})$.}
At early times, the axial length scale is $L_x \sim \sqrt{t}$ and molecular diffusion dominates. 
The convective and axial-diffusive terms scale as
\begin{equation}
	\Pen u\frac{\partial C}{\partial x}
	\sim\frac{\Pen}{\sqrt{t}},
	\qquad
	\frac{\partial^2C}{\partial x^2}
	\sim\frac{1}{t},
\end{equation}
so that the diffusive regime requires
\begin{eqnarray}
	\frac{\Pen}{\sqrt{t}} \ll \frac{1}{t}
	\quad\Longrightarrow\quad
	t \ll t_c^{o} \sim \Pen^{-2},
\end{eqnarray}
For $\Pen=100$, this gives $t_c^{o}\sim10^{-4}$. The leading-order variance for a central point source is
\begin{eqnarray}
	\bar{\sigma}^2 \propto 2t,
	\label{simga_initial}
\end{eqnarray}
and hence $\alpha \simeq 1$.

\subsubsection{Regime \RomanNum{2}: Anomalous and pre-asymptotic regimes ($t_c^{o} \ll t \ll t_c^{d} $).}
The subsequent evolution is controlled by the coupled effects of radial diffusion, velocity shear and interphase mass transfer.
For an inert solute, the transient increase of $\alpha$ reflects the progressive conversion of radial velocity heterogeneity into streamwise spreading as solute samples the non-uniform velocity profile.
For the reactive problem, the behaviour depends on the reactivity, and we therefore distinguish small and large Damköhler numbers.

\emph{Reaction-limited regime:} $k_s\ll O(1)$.
When the reaction is weak, shear-enhanced spreading remains the dominant mechanism in the intermediate regime.
For the tube Poiseuille profile and a central point source ($r_0\to0$), the short-time expansion of the axial variance is
\begin{eqnarray}
	\bar{\sigma}^2 \propto 2t+ \frac{32}{3}\Pen^2 t^4 + \frac{32}{3}\Pen^2 r_0^2 t^3 = 2t+ \frac{32}{3}\Pen^2 t^4.
	\label{simga_pre}
\end{eqnarray}
The absence of a $t^3$ contribution for $r_0=0$ follows from the vanishing initial radial distance from the centreline. 
Balancing the molecular and shear-induced contributions gives the characteristic scaling
$	t \sim \Pen^2 t^4,$
and hence provides an estimate of the onset of the shear-dominated transient regime
\begin{eqnarray}
	t_c^b \sim \Pen^{-2/3}.
\end{eqnarray}
which gives $t_c^b\simeq 0.05$ for $\Pen=100$ at the level of scaling estimates, in good agreement with the time at which $\alpha$ reaches the unimodal peak and depart appreciably from unity in figure \ref{fig_scaling}($b$). 
Thus, for weak reaction, the anomalous excursion of $\alpha$ remains primarily a shear-induced pre-asymptotic effect, with wall reaction providing a perturbative correction.

\emph{Diffusion-limited regime}: $k_s\gg O(1)$.
For stronger reaction, the transient evolution is governed by the competition between shear-induced streamwise transport and relaxation of the radial concentration field under the Robin boundary condition. 
A single reaction time proportional to $1/k_s$ is therefore insufficient to describe the observed non-monotonic evolution.

The relevant radial timescales follow from the eigenvalue spectrum. For $k_s=0$, $\lambda_0=0$ expresses conservation of total mass and the first non-zero eigenvalue gives the radial mixing scale.
For $k_s>0$, the lowest eigenvalue $\lambda_0(k_s)>0$ instead represents the slow decay of the total mass. 
It approaches the lowest Dirichlet eigenvalue,
$\lim\limits_{k_s\to\infty}\lambda_0 \approx 5.783,$
whereas the next eigenvalue governs the slowest relaxation of the cross-sectional concentration profile once the mass-loss mode has been excluded. 
Accordingly, we define
\begin{equation}
	t_c^m=\frac{1}{\lambda_0(k_s)},
	\qquad
	t_c^r(k_s)=\frac{1}{\lambda_1(k_s)}.
\end{equation}
For large $k_s$, $\lim\limits_{k_s\to\infty}t_c^m \approx 0.173,$
while $\lim\limits_{k_s\to\infty}t_c^r	\approx0.033.$
For finite $k_s$, these timescales remain distinct: $t_c^m$ describes overall mass loss, whereas $t_c^r$ characterizes cross-sectional profile relaxation. 
They provide natural time scales for the troughs and peaks observed in $\alpha$, although the locations of those extrema need not coincide exactly with the corresponding inverse eigenvalues. Increasing $k_s\to\infty$ accelerates both processes until the radial relaxation becomes limited by radial diffusion.

\subsubsection{Regime \RomanNum{3}: Taylor dispersion regime ($t \gg t_c^d \sim 1$).}

At sufficiently long times, radial diffusion has sampled the full tube cross-section, with the radial diffusion length satisfying $\sqrt{t_c^d}\sim1$. Hence
\begin{eqnarray}
	t \gg t_c^d \sim 1.
\end{eqnarray}
The recovery of $\alpha \to 1$ signals the Taylor dispersion regime, in which the cross-sectional distribution has relaxed and the axial variance grows linearly with time,
\begin{eqnarray}
	\bar{\sigma}^2 \propto 2\bar{D}_\text{eff} t + \text{const}.
	\label{simga_taylor}
\end{eqnarray}

The present results extend a broader body of work from inert solutes to chemically active solutes and, importantly, identifying the intermediate regimes through the scaling relation $\bar{\sigma}^2\propto t^\alpha$.
\cite{taylorDispersionSolubleMatter1953} and \cite{arisDispersionSoluteFluid1956} established the long-time effective-diffusion description, while \cite{lighthillInitialDevelopmentDiffusion1966} analyzed the early development of dispersion and identified the importance of the initial shear-induced head. 
\cite{chatwinApproachNormalityConcentration1970} further examined the approach towards the asymptotic state, and \cite{phillipsInitialTransientConcentration1997} developed asymptotic descriptions of the initial transient. 
The review of \cite{youngShearDispersion1991} emphasized that the one-dimensional Taylor equation is fundamentally a long-time reduction and that anomalous, non-Fickian spreading may persist well into the pre-asymptotic regime.

\section{Streamwise dispersion with interphase mass transfer: asymptotics and patterns}
\label{result}

The concentration field represented by (\ref{con_streamwiseExpand_sol}) can be regarded rigorously as the solution of a generalized dispersion equation, expressed directly in terms of spatial derivatives of all orders:
\begin{eqnarray}
	\frac{\partial C}{\partial t} = \sum\limits_{n=0}^{\infty} 
	K_n(r,t) \frac{\partial^n C}{\partial x^n},\quad n=0,1,2,\ldots,
	\label{general_dispersion}
\end{eqnarray}
where $K_n$ are the streamwise transport coefficients defined by the time evolution of the local cumulants,
\begin{eqnarray}
	K_n = \frac{(-1)^n}{n!} \frac{\partial \kappa_n}{\partial t},
\end{eqnarray}
where $\kappa_n$ are the local cumulants introduced in (\ref{con_chagen_def}), with the first five given explicitly in the preceding analysis.
Note that the long- and short-time asymptotic solutions of (\ref{general_dispersion}) matches (\ref{eq_long_meanC}) and (\ref{sol_guan_particular}).
We truncate this expansion at orders $n=2,3,4$, and denote them as the second-, third-, and fourth-order streamwise models, respectively.

This link between the hierarchy of spatial derivatives and the statistics of the concentration distribution is a distinctive feature of the present theory: the entire non‑Gaussian evolution is captured by the sequence of coefficients $K_n$, each of which can be computed from the analytical solution for the moments.
Specifically, $K^S$ is the local kinetic exchange rate due to chemical reactions:
\begin{eqnarray}
	K^S \triangleq \frac{\mathrm{d}}{\mathrm{d} t} \ln C_0 = K_0,
\end{eqnarray}
which quantifies the exponential decay of the solute due to the wall reaction at each radial position.
It is negative for finite $k_s$ at long times and governs the overall mass loss.

Higher order $K_n$ are associated with the streamwise transport coefficients.
For example, the streamwise drift $U^S$ relates to $K_1$ through
\begin{eqnarray}
	U^S \triangleq \frac{\partial \kappa_1}{\partial t} = -K_1.
\end{eqnarray}
This local drift velocity captures how the centre of mass of the solute cloud moves along each streamline.
It incorporates both the convective motion of the fluid and any bias introduced by the interaction of shear with radial diffusion and surface reaction.

Likewise, streamwise dispersivity $D^S$ equals $K_2$ via
\begin{eqnarray}
	D^S \triangleq \frac{1}{2}\frac{\partial \kappa_2}{\partial t} = K_2,
\end{eqnarray} 
which is the instantaneous rate of growth of the streamwise variance, i.e., the local dispersivity.
At long times, after radial averaging, it reduces to the classical Taylor--Gill dispersivity, but here it retains its full time and radial dependence.

\begin{figure}
	\centerline{\includegraphics{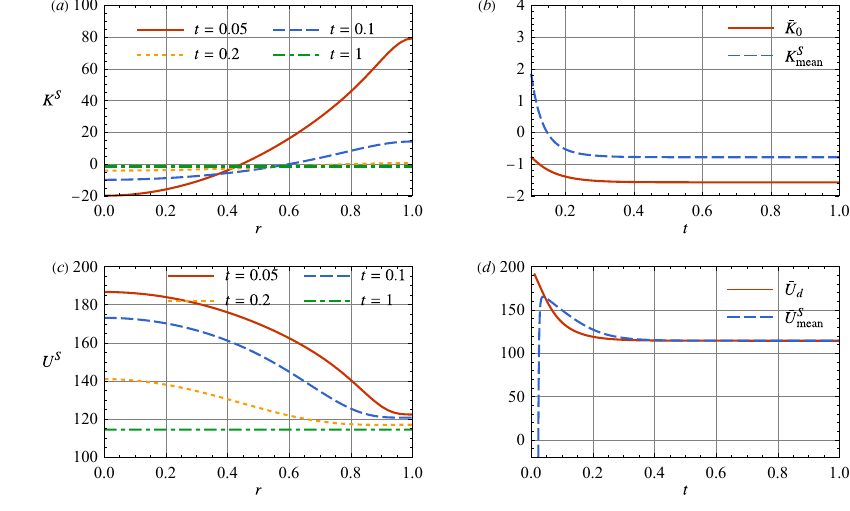}}
	\caption{Variation of streamwise exchange rate and drift velocity with $r$ and $t$ for $k_s=1$ and $\Pen=100$.}
	\label{fig_coe_K0US}
\end{figure}

Figure \ref{fig_coe_K0US}($a$) shows radial--temporal structure of the local exchange rate. 
In contrast to the uniformly negative profile of $\bar{K}_0 \triangleq \partial_t \ln M_0$, $K^S$ reveal a sign-changing radial distribution at early times.
At $t=0.05$, $K^S$ is strongly negative in the core, and reaches approximately $80$ near the wall.
This indicates that the local tendency of the concentration field is highly non-uniform: the core experiences a local depletion, whereas the near-wall region exhibits a positive local exchange rate.
The reaction consumes solute near the wall, radial diffusion transports solute from the core towards the wall, and the resulting radial redistribution produces opposite signs of $K^S$ in different regions.
As $t$ increases to $0.1$, the magnitude of $K^S$ decreases substantially and the radial contrast weakens.
By $t=1$, $K^S$  is nearly uniform and close to zero across the entire cross-section. This suggests that the wall reaction becomes limited by cross-sectional mixing and that the local exchange rate approaches a quasi-steady, radially homogenized state. 
The behaviour is consistent with the decay of the mean exchange rate shown in figure \ref{fig_coe_K0US}($b$).

The corresponding evolution of local drift $U^S$ and global drift $\bar{U}_d \triangleq \partial_t\bar{\mu}$, with $\bar{\mu}\triangleq M_1/M_0$, are shown in figure \ref{fig_coe_K0US}($c,d$). 
At early times, $U^S$ decreases systematically with $r$, reflecting the underlying Poiseuille shear: solute near the centreline samples a larger streamwise velocity than solute near the wall. 
Its temporal evolution is therefore governed not simply by a local modification of the velocity field, but by the radial redistribution of the concentration. 
As reaction proceeds, the near-wall population is preferentially removed, shifting the surviving solute towards the faster core streamlines. 
The concentration-weighted mean velocity consequently increases, while the radial variation of $U^S$ diminishes as radial mixing proceeds. 
This redistribution ultimately leads to the common long-time value of the mean drift shown in figure \ref{fig_coe_K0US}($d$).

\begin{figure}
	\centerline{\includegraphics{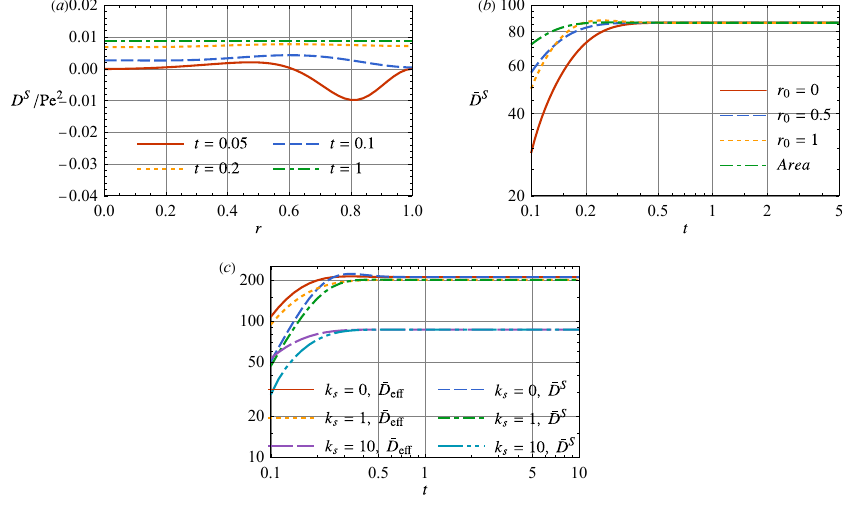}}
	\caption{Evolution of streamwise dispersivity for $k_s=10$ and $\Pen=100$: ($a$) variation of streamwise dispersivity normalized by $\Pen^2$ with $r$; ($b$) variation of mean streamwise dispersivity with $t$ for different initial conditions; and ($c$) variation of mean streamwise dispersivity and mean Taylor dispersivity, with inert solutes recovering the classical Taylor--Aris limits $1+\Pen^2/48$.}
	\label{fig_coe_DT}
\end{figure}

Figure \ref{fig_coe_DT}($a$) demonstrates that $D^S$ is strongly dependent on both $r$ and $t$, particularly during the transitional stage. 
Some local curves become negative at intermediate times. 
This should not be interpreted as a negative molecular or effective diffusivity.
Rather, a negative value merely indicates that the local second cumulant is temporarily decreasing as radial redistribution changes the population of streamwise trajectories. 
Such non-monotonic behaviour is therefore a natural consequence of describing a non-equilibrated reactive population through local cumulants.

The radially averaged dispersivity in figure \ref{fig_coe_DT}($b$) exhibits a rapid transient followed, for the different initial conditions, by convergence towards an approximately common plateau. 
The initial condition primarily determines how rapidly the radial distribution approaches the reactive eigenmode, after which the system loses memory of the initial radial localization. 
A modest overshoot for some initial conditions is consistent with the transient interaction between shear-induced dispersion and cross-sectional mixing. 

Once the quasi-equilibrium state is established, $D^S$ becomes nearly constant.
This behaviour has a direct implication for the variance, where $\kappa_2 \simeq 2 D^S t +\text{const}$ at long times.
The streamwise variance ultimately grows diffusively, while the early-time evolution is transitional because $D^S$ itself varies substantially with time.

Figure \ref{fig_coe_DT}($c$) compares the mean streamwise dispersivity with the corresponding Taylor dispersivity $\bar{D}_\text{eff}\triangleq \partial_t{\bar{\sigma}^2}/2$. 
Both quantities approach a common long-time value of order $\Pen^2$, demonstrating recovery of the classical Taylor–Aris scaling once cross-sectional mixing has equilibrated.
For $\Pen=100$ and $k_s=0$, the plateau is $1+\Pen^2/48$, consistent with the familiar results of inert solutes in tube Poiseuille flows.
The transient difference between $\bar{D}_\text{eff}$ and $\bar{D}^S$, including the overshoot of the latter, reflects the fact that the mean Taylor--Aris dispersivity is an asymptotic quantity.
More generally, increasing reactivity reduces the asymptotic dispersivity relative to the non-reactive case because the catalytic wall preferentially removes solute from regions that participate strongly in the cross-stream velocity contrast. 
The resulting concentration profile has reduced shear-weighted variance, while the altered residence-time distribution further suppresses the contribution of slow near-wall trajectories to shear-induced spreading. 
Thus, although the long-time scaling remains $\displaystyle D^S \sim \mathit{O}(\Pen^2)$, the coefficient multiplying $\Pen^2$ becomes dependent on the reactivity, or equivalently on the Damköhler number. 
In this sense, interphase mass transfer modifies the population of trajectories from which the Taylor mechanism is generated.

\subsection{Local transport coefficients: small- and large-$k_s$ asymptotics}

Recall that the eigenvalues are determined by the roots of $\displaystyle J_0(\alpha) k_s={\alpha J_1(\alpha)}$, with the smallest positive root denoted by $\alpha_0$. 
For $k_s>0$, all decay rates are therefore positive and the slowest mode decays as $\displaystyle \exp({-\alpha_0^2t})$.
The fundamental eigenvalues provide a compact interpolation between two distinct physical mechanisms: reaction-limited decay for small $k_s$ and diffusion-limited decay approaching the Dirichlet condition for large $k_s$.
As $k_s\to0$, the solution converges to the inert case with a no-penetration condition at the wall \citep{guanStreamwiseDispersionSoluble2024}.
As $k_s$ increases, the concentration near the surface is increasingly depleted; in the $k_s \to \infty$, the boundary condition becomes Dirichlet, analogous to the constant-wall-temperature case in heat transfer.
Below, useful asymptotic results are derived for the small- and large-Damköhler-number limits.

\subsubsection{Limiting cases of exchange rate}
The local and global exchange rates are given by
\begin{eqnarray}
	K^S(r,t) = -\frac{ \displaystyle\sum_i \alpha_i^2A_iJ_0(\alpha_i r)e^{-\alpha_i^2t}} {\displaystyle\sum_i A_iJ_0(\alpha_i r)e^{-\alpha_i^2t}},\quad 
	\bar K_0(t) = -\frac{ \displaystyle\sum_i \alpha_i^2B_i e^{-\alpha_i^2t}} {\displaystyle\sum_i B_i e^{-\alpha_i^2t}}.
\end{eqnarray}
with
\begin{eqnarray}
	A_i= \frac{\alpha_i^2} {(k_s^2+\alpha_i^2)J_0^2(\alpha_i)},\quad B_i= \frac{2k_s} {(k_s^2+\alpha_i^2)J_0(\alpha_i)}.
\end{eqnarray}
Both quantities can therefore be interpreted as weighted averages of the modal decay rates $-\alpha_i^2$.

For the central-point release, the source has not yet reached the wall at short times.
Thus, $\lim\limits_{t\to0}K^S \sim 0$, with the $k_s$-induced correction exponentially small, while the initial diffusion transient $\sim \exp[{-1/(4t)}]$ dominates.
Using the $k_s$-perturbation of the Bessel expansions,
the long-time exchange rate in the small $k_s$ limit is
\begin{eqnarray}
	\displaystyle 
	\lim\limits_{t\to\infty}K^S \sim \lim\limits_{t\to\infty}\bar{K}_0 \sim -\lambda_0 = -2k_s + \frac{1}{12} k_s^2 - \frac{1}{12} k_s^3 +\mathit{O}(k_s^4),\quad k_s \ll 1,
	\label{K0_small_ks}
\end{eqnarray}
whereas in the large-$k_s$ limit,
\begin{eqnarray}
	\displaystyle 
	\lim\limits_{t\to\infty}K^S \sim \lim\limits_{t\to\infty}\bar{K}_0 \sim K^S \sim -\lambda_0+\frac{2\lambda_0}{k_s} +O(k_s^{-2}),\quad k_s \gg 1.
	\label{K0_large_ks}
\end{eqnarray}
with $ \lim\limits_{k_s\to\infty}\lambda_0 \approx 5.783$.
At long times, both local and global exchange rates are governed by the single slowest mode, $\lambda_0$, and hence decay exponentially.
During the transitional regime, several radial eigenmodes contribute simultaneously.
Key results for the local and global exchange rates are summarized in table \ref{tab:exchange_summary}.

\begin{table}
	\begin{center}
		\def~{\hphantom{0}}
		\setlength{\tabcolsep}{4pt}
		\begin{tabular}{p{0.14\linewidth} p{0.2\linewidth} p{0.25\linewidth} p{0.3\linewidth}}
			Regime
			&
			Characteristic behaviour
			&
			Exchange rate
			&
			Physical interpretation
			\\[6mm]
			$t\ll1$
			& Central source has not reached the wall
			& $\displaystyle \bar{K}_0(t)\rightarrow0$; wall correction exponentially small,
			& Diffusion dominates; wall reaction not activated appreciably.
			\\[2mm]		
			$t\rightarrow\infty$
			& $\lambda_0$ dominates
			& $\displaystyle K^S(r,t)\sim\bar{K}_0(t)\sim-\lambda_0$
			& Single-mode exponential decay.
			\\[2mm]
			$k_s\ll1$
			& Weak reaction
			& $\displaystyle K^S \sim -2k_s+\frac{1}{12} k_s^2+O(k_s^3)$
			& Reaction-limited; rapid mixing; approach to radial uniformity.
			\\[2mm]
			$k_s\gg1$
			& Strong reaction
			& $\displaystyle K^S \sim -\lambda_0+\frac{2\lambda_0}{k_s} +O(k_s^{-2})$
			& Diffusion-limited; approach to a perfectly absorbing boundary.
			\\[2mm]
			General \newline mechanism
			& Several radial eigenmodes contribute
			& $\displaystyle K^S=\partial_t C_0,~\bar{K}_0 = \partial_t M_0$; weighted averages of $-\lambda_i$
			& Radial redistribution and surface reaction occur simultaneously.
			\\
		\end{tabular}
		\caption{Asymptotic behaviour of the exchange rate, with small- and large-$k_s$ limits.}
		\label{tab:exchange_summary}
	\end{center}
\end{table}

\subsubsection{Limiting cases of drift}

As $t\to0$, the local and global drifts for a uniform area source are
\begin{eqnarray}
	U^S(r,t) \sim \Pen u(r)+ \mathit{O}(t^{1/2}), \quad \bar{U}_d(t)=\Pen \bar u+\mathit{O}(t^{1/2}).
\end{eqnarray}
Hence, to leading order in time, the initial local and global drifts follow the local and mean flow velocities, respectively.
This is in accordance with the particular solution (\ref{sol_guan_particular}) at asymptotically short times in a free space.

Likewise, for a central point source,
\begin{eqnarray}
	U^S(0,t) \sim \Pen u(0)+ \mathit{O}(t).
\end{eqnarray}
Thus, the initial local drift is set by the local velocity at the source location, rather than by the mean flow velocity, see (\ref{eq_short_meanC}).

The long-time local drift is given by
\begin{eqnarray}
	U^S = \frac{\partial\kappa_1}{\partial t} = \Pen \overline{u f_0^2} + \mathit O\!\left(te^{-(\lambda_1-\lambda_0)t}\right).
\end{eqnarray}
Thus, at long times, the local and global drifts have the same leading coefficient:
\begin{eqnarray}
    \lim\limits_{t\to\infty}U^S =\lim\limits_{t\to\infty} \bar U_d \sim \Pen\,\overline{u f_0^2}.
\end{eqnarray}
For small $k_s$, the local drift approaches
\begin{eqnarray}
	\displaystyle 
	\lim\limits_{t\to\infty} U^S \sim \Pen(\bar{u}+\frac{k_s}{6}) +\mathit{O}(k_s^2).
	\label{Ud_small_ks}
\end{eqnarray}
That is, the drift approaches the mean flow velocity at leading order in $k_s$, as reflected in (\ref{eq_long_meanC}).
The physical origin of this behaviour is that the principal eigenfunction becomes spatially uniform before the very slow reaction takes place. 
Consequently, the first-mode weighting $f_0^2$ becomes equivalent to the uniform cross-sectional weighting.

For large $k_s$ the drift becomes
\begin{eqnarray}
	\displaystyle 
	\lim\limits_{t\to\infty} U^S \sim \frac{4\Pen(\lambda_0+1)}{3\lambda_0}
	-\frac{4\Pen(\lambda_0-1)}{3\lambda_0 k_s}
	+O(k_s^{-2})
	\approx 1.56\Pen-1.1\Pen k_s^{-1}+O(k_s^{-2}).
	\label{Ud_large_ks}
\end{eqnarray}
The underlying mechanism is that reaction suppresses the concentration near the wall, while the drift is weighted by the square of the dominant radial eigenfunction, leading to centreline-biased transport.

Therefore, weak reaction leaves the principal mode nearly uniform and recovers the bulk mean velocity, whereas strong reaction drives the mode towards the first Dirichlet Bessel mode, producing centreline-biased transport and a diffusion-limited decay rate.
Key results for the local and global exchange rates are summarized in table \ref{tab:drift_summary}.

\begin{table}
	\begin{center}
		\def~{\hphantom{0}}
		\setlength{\tabcolsep}{4pt}
	   \begin{tabular}{p{0.12\linewidth} p{0.2\linewidth} p{0.25\linewidth} p{0.3\linewidth}}
		Regime
		&
		Characteristic behaviour
		&
		Drift
		&
		Physical interpretation
		\\[6mm]
		$t\to0$,
		&
		Uniform source
		&
		$ U^S \sim \Pen u(r)$, while
		$ \bar U_d = \Pen\bar u$
		&
		Diffusion transient; drifts follow the local and mean velocities.
		\\[2mm]
		$t\to0$
		&
		Central point source
		&
		$U^S\sim \Pen u(0)$, while $\bar U_d \sim \Pen u(0)$
		&
		In the fast core; drift set by centreline velocity.
		\\[2mm]
		$t\to\infty$, \newline $k_s\ll1$
		&
		Weak reaction
		&
		$\displaystyle U^S \sim \Pen\bar u+\Pen k_s/6+\mathit{O}(k_s^2)$
		&
		Reaction-limited; drift approaches mean velocity.
		\\[2mm]
		$t\to\infty$,\newline $k_s\gg1$
		&
		Strong reaction
		&
		$\displaystyle 
		U^S \sim \Pen\overline{uf_0^2}$
		$\displaystyle \approx 1.56\Pen-1.1\Pen k_s^{-1}+O(k_s^{-2})$
		&
		Diffusion-limited; $\overline{u f_0^2}={4(\lambda_0+1)}/{(3\lambda_0)}$.
		\\[2mm]
		General mechanism
		&
		Reaction changes several radial modes
		&
		$\displaystyle U^S = \partial_t \kappa_1, \bar{U}_d = \partial_t \bar{\mu}$; weighted averages of $\lambda_i$
		&
		From the classical bulk mean to the Dirichlet Bessel mode.
		\\
	\end{tabular}
		\caption{Asymptotic behaviour of the drift, with small- and large-$k_s$ limits.}
	\label{tab:drift_summary}
    \end{center}
\end{table}

\subsubsection{Limiting cases of dispersivity}

For asymptotically long times, a convenient spectral representation for streamwise dispersivity is
\begin{eqnarray}
	\lim\limits_{t\to\infty}D^S = 1 +\Pen^2 \sum_{i=1}^{\infty} \frac{ (\overline{ u f_0 f_i})^2 }{ \lambda_i-\lambda_0 },
\end{eqnarray}
which is a reactive analogue of the Taylor dispersivity.
For $k_s\ll1$,
\begin{eqnarray}
	D^S \sim 1+\Pen^2 \left[ \frac1{48} +\frac{k_s}{1440} -\frac{37k_s^2}{17280} +O(k_s^3) \right].
	\label{DS_small_ks}
\end{eqnarray}
In particular, the conservative Taylor--Aris limit is recovered smoothly: $\displaystyle D^S(k_s\to0) = 1+{\Pen^2}/{48}.$

For $k_s\gg1$, the local dispersivity becomes
\begin{eqnarray}
	\lim\limits_{t\to\infty} D^S = 1+\Pen^2 \left[ \mathcal S_0 +\frac{6\mathcal S_0}{k_s} +O(k_s^{-2}) \right]
	\approx 1+0.005\,\Pen^2+\frac{0.03\Pen^2}{k_s}.
	\label{DS_large_ks}
\end{eqnarray}
with
\begin{eqnarray}
	\mathcal S_0 = \frac{4(-\lambda_0^2+17\lambda_0-54)}{45\lambda_0^3} < \frac{1}{48}.
\end{eqnarray}
Thus, the Dirichlet boundary condition gives a substantially smaller long-time dispersivity than a conservative wall.
This behaviour has a natural physical interpretation. 
In the $k_s\to\infty$ limit the solute satisfies the Dirichlet condition and is strongly depleted near the wall. 
The surviving concentration is therefore concentrated towards the tube centre, where the Poiseuille shear is relatively small. 
The velocity contrast sampled by the surviving solute is consequently reduced, suppressing streamwise dispersion.

For large but finite $k_s$, the Robin condition permits a small but $O(k_s^{-1})$ wall concentration. 
The solute therefore samples more of the low-speed near-wall region, increasing the velocity contrast and hence increasing $D^S$. 
The positive coefficient $6 \mathcal S_0$ is precisely this first finite-boundary correction.
Key results for the local and global dispersivities are summarized in table \ref{tab:dispersivity_summary}.

\begin{table}
	\begin{center}
		\setlength{\tabcolsep}{4pt}
		\begin{tabular}{p{0.14\linewidth} p{0.2\linewidth} p{0.28\linewidth} p{0.28\linewidth}}
			Regime & Characteristic behaviour & Dispersivity & Physical interpretation \\[6mm]
			$t\to\infty$,\newline$\ k_s\ll1$ &
			Reaction limited &
			$\displaystyle D^S \sim 1+ \frac{\Pen^2}{48} +\frac{\Pen^2 k_s}{1440}$
			&
			Reproduces classical results with small $k_s$ corrections. \\[6mm]
			$t\to\infty$,\newline $k_s\gg1$ &
			Diffusion limited &
			$ \displaystyle D^S \sim 1+0.005\,\Pen^2+\frac{0.03\Pen^2}{k_s}$
			&
			Converges to the Dirichlet condition; reaction dominates. \\[6mm]
			General \newline mechanism &
			Competition between diffusion, shear and reaction &
			$\displaystyle D^S=\partial_t \kappa_2$
			with transient non-monotonic behaviour &
			Reaction removes high-shear population and changes pre-asymptotic dispersion. \\[6mm]
	  \end{tabular}
		\caption{Asymptotic behaviour of the dispersivity, with small- and large-$k_s$ limits.}
		\label{tab:dispersivity_summary}
	\end{center}
\end{table}

\begin{figure}
	\centerline{\includegraphics{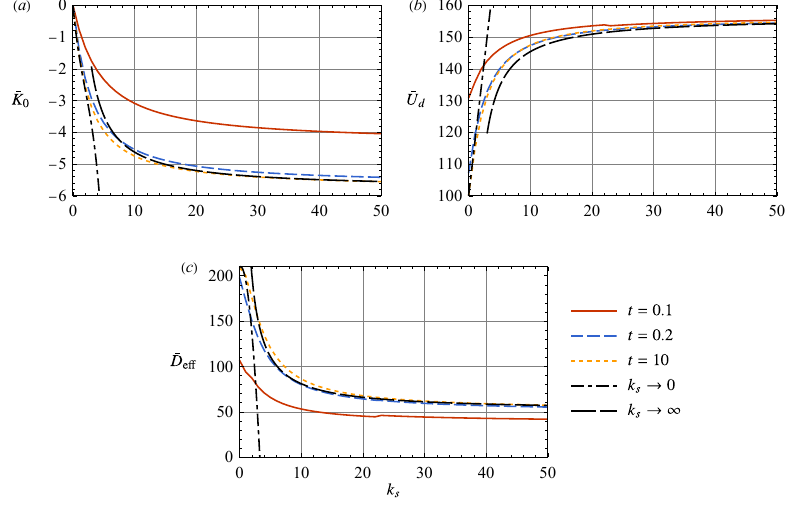}}
	\caption{Variation of the global quantities with the Damköhler number for a central point source: ($a$) mean exchange rate $\bar{K}_0$, ($b$) mean drift, and ($c$) mean dispersivity for $\Pen=100$. The black lines are small- and large-$k_s$ asymptotics given by equations (\ref{K0_small_ks}), (\ref{K0_large_ks}), (\ref{Ud_small_ks}), (\ref{Ud_large_ks}), (\ref{DS_small_ks}), and (\ref{DS_large_ks}), in good agreement with the general analytical results.}
	\label{fig_coe_k0}
\end{figure}

The radial redistribution induced by interphase mass transfer also provides a natural explanation for the trends in the global quantities. 
In the asymptotic long-time regime, the global moments coincide with their local counterparts, so the small- and large-$k_s$ asymptotic results derived above can be used directly to interpret the mean exchange rate, mean drift, and mean dispersivity in figure \ref{fig_coe_k0}.
The evolution of $\bar{K}_0$ becomes increasingly negative as $k_s$ becomes large, evolving from $\bar{K}_0\to 0$ in the reaction-limited regime to the Dirichlet limit $-\lambda_0\approx -5.783$.
The black asymptotic curves obtained from equations (\ref{K0_small_ks}) and (\ref{K0_large_ks}) capture these limiting behaviours well.

The mean drift $\bar{U}_d$ increases with $k_s$, as shown in figure \ref{fig_coe_k0}($b$). 
Wall reaction preferentially removes solute from the near-wall, low-velocity region, shifting the surviving concentration towards the faster core streamlines. 
The concentration-weighted mean drift therefore increases with $k_s$, consistent with the small- and large-$k_s$ asymptotic results given by equations (\ref{Ud_small_ks}) and (\ref{Ud_large_ks}).

In contrast, $\bar{D}_\text{eff}$ decreases with $k_s$, as shown in figure \ref{fig_coe_k0}($c$). 
The preferential depletion of near-wall solute weakens the contribution of the high-shear region and reduces the radial concentration gradients responsible for Taylor dispersion. Consequently, the long-time dispersivity slightly increases from the conservative limit and soon decreases to a lower asymptotic value as $k_s$ increases, in agreement with equations (\ref{DS_small_ks}) and (\ref{DS_large_ks}). 
Thus, the global trends in drift and dispersion follow directly from the reaction-induced radial redistribution of the surviving solute.

\subsection{Concentration patterns: beyond classical theories}

\begin{figure}
	\centerline{\includegraphics{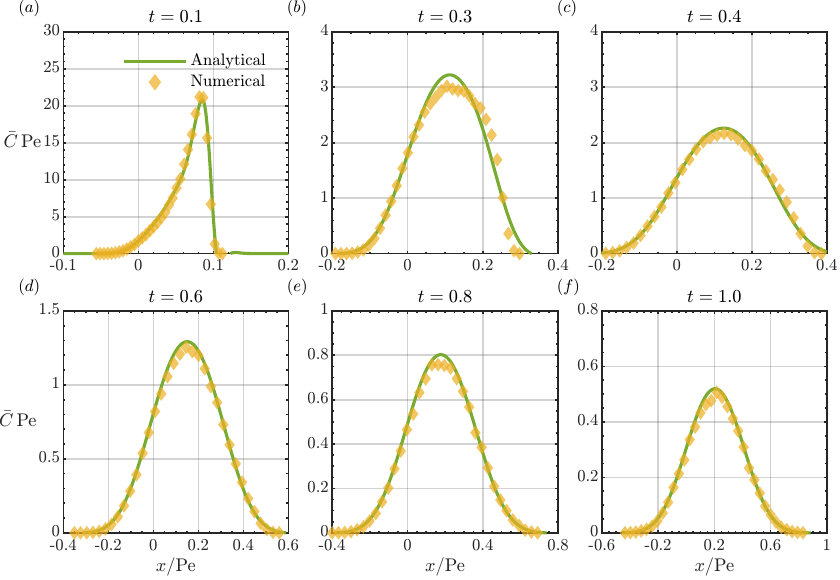}}
	\caption{Validation of fourth-order analytical solutions for mean concentration (lines) against direct Monte Carlo simulations (diamonds) for the mean concentration. The solute is initially released from a point source at the centre. The parameters are $k_s=1$, $\Pen=100$, $N=10^{6}$, and $\Delta t=10^{-4}$.}
	\label{fig_con_pt}
\end{figure}

The analytical solution is validated against direct Monte Carlo simulations. 
As shown in figure \ref{fig_con_pt}, the predicted mean concentration exhibits quantitative agreement with the simulations across all the regimes considered.

\begin{figure}
	\centerline{\includegraphics{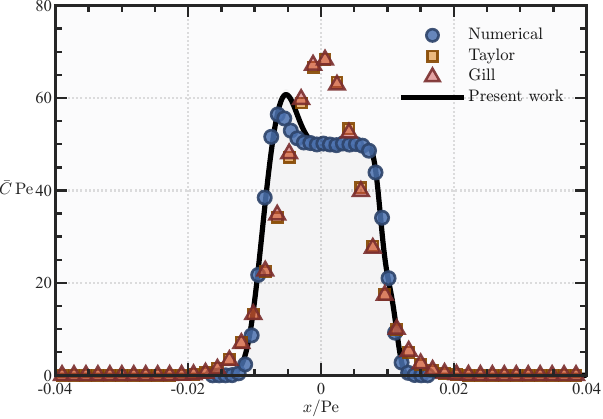}}
	\caption{Mean concentration derived from analytical solutions of the present second-order model (line) validated against direct Monte Carlo simulations (circle), compared with classical models \citep[see e.g.][]{taylorDispersionSolubleMatter1953,gillExactAnalysisUnsteady1970,sankarasubramanianUnsteadyConvectiveDiffusion1973}. The solute is initially released from a uniform area source. The parameters are $t=0.01$, $k_s=1$, $\Pen=100$, $N=10^{7}$, and $\Delta t=10^{-5}$.}
	\label{fig_mean_con_area_t0.01}
\end{figure}

\begin{figure}
	\centerline{\includegraphics{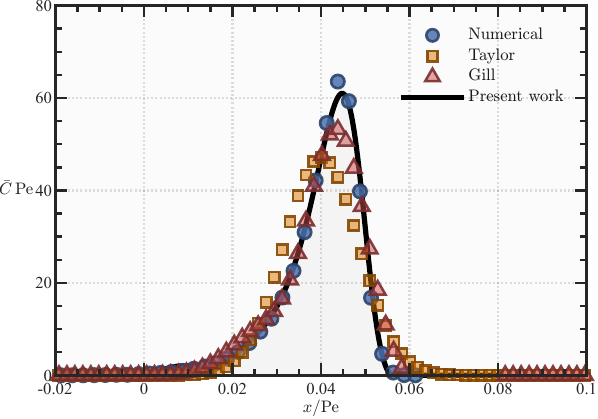}}
	\caption{Mean concentration derived from analytical solutions of the present third-order model (line) validated against direct Monte Carlo simulations (circle), compared with classical models \citep[see e.g.][]{taylorDispersionSolubleMatter1953,gillExactAnalysisUnsteady1970,sankarasubramanianUnsteadyConvectiveDiffusion1973}. The solute is initially released from a central point source. The parameters are $t=0.05$, $k_s=1$, $\Pen=100$, $N=10^{7}$, and $\Delta t=10^{-5}$.}
	\label{fig_mean_con_area_t0.05}
\end{figure}

Figures \ref{fig_mean_con_area_t0.01} and \ref{fig_mean_con_area_t0.05} further elucidate the performance of the present streamwise model by comparing the analytical solution with numerical simulations and classical dispersion models. 
For the uniform area source at $t=0.01$ (figure \ref{fig_mean_con_area_t0.01}), the present streamwise model reproduces the full concentration profile, including the pronounced departure from a single Gaussian peak, whereas the classical approximations progressively lose accuracy during the transitional regime. 
This behaviour is closely reminiscent of the non-Gaussian structure reported by \cite{smithGAUSSIANAPPROXIMATIONCONTAMINANT1982} and \cite{houseworthShearDispersionResidence1984}, who showed that a Gaussian approximation could not recover the two-peaked exact profile and introduced an ad hoc skewness correction, with the Gaussian description remaining useful only over approximately one $e$-folding time. 
The present formulation instead captures the skewness intrinsically through the second-order streamwise model, without requiring an empirical correction. 

A closely related correspondence is observed in figure \ref{fig_mean_con_area_t0.05} for a central point source at $t=0.05$, where the analytical profile closely follows the Monte Carlo results and recovers the asymmetric, distinctly non-Gaussian tails also highlighted by \cite{latiniTransientAnomalousDiffusion2001}. 
In contrast to the separate head and tail approximations adopted by Latini \& Bernoff, the present solution provides a compact representation over the whole profile without introducing separate asymptotic expressions for the leading and trailing regions. 
Moreover, whereas both \cite{smithGAUSSIANAPPROXIMATIONCONTAMINANT1982} and \cite{latiniTransientAnomalousDiffusion2001} considered the non-reactive case $k_s=0$, the present formulation considers the wall-reaction term and therefore captures the modification of the pre-asymptotic concentration field for $k_s>0$.

\begin{figure}
	\centerline{\includegraphics{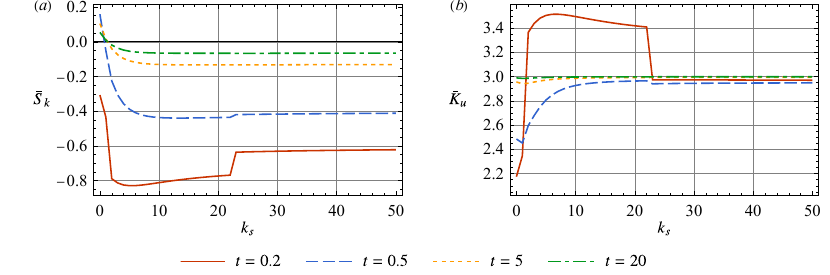}}
	\caption{Variation of dispersion coefficients with different Damköhler numbers: ($a$) mean skewness $\bar{S}_k$, and ($b$) mean kurtosis  $\bar{K}_u$ for $\Pen=100$.}
	\label{fig_coe_sk}
\end{figure}

The statistical origin of these non-Gaussian features is further clarified by figure \ref{fig_coe_sk}, which shows the dependence of the mean skewness $\bar S_k \triangleq ({M_3}/{M_0} - 3\bar\mu M_2/M_0 +2 \bar\mu^3)/\bar{\sigma}^3$ and mean kurtosis $\bar K_u \triangleq ({M_4}/{M_0} - 4\bar\mu{M_3}/{M_0} - 3 M_2/M_0 + 12\bar\mu^2 {M_2}/{M_0} - 6\bar\mu^4)/ \bar{\sigma}^4$ on the reactivity $k_s$ and time $t$. 
During the transitional regime, $\bar S_k$ departs appreciably from its Gaussian reference value, indicating a finite asymmetry of the streamwise concentration distribution, while the concomitant evolution of $\bar K_u$ signals a substantial departure from Gaussian peakedness. 
The magnitudes of these higher-order moments vary systematically with both $t$ and $k_s$, demonstrating that wall reaction does not merely rescale the concentration amplitude but alters the shape of the evolving distribution. 
In particular, the persistence of non-zero $\bar S_k$ together with the corresponding deviation of $\bar K_u$ from its Gaussian reference under $k_s \neq 0$ provides a statistical measure of the asymmetric and double-peaked profiles observed in figures \ref{fig_mean_con_area_t0.01} and \ref{fig_mean_con_area_t0.05}. 
Thus, the emergence and subsequent attenuation of the non-Gaussian concentration can be interpreted directly in terms of the temporal evolution of the third- and fourth-order moments: skewness governs the unequal development of the upstream and downstream portions of the profile, while kurtosis quantifies the enhanced concentration peaking associated with the transitional state. 
The agreement among the full concentration fields, the Monte Carlo simulations and the higher-order moment evolution shows that the streamwise formulation inherently captures the salient non-Gaussian features that previously required either an empirical skewness correction or separate head–tail approximations, while simultaneously extending the description to arbitrary reactivity.

\section{Conclusions}
\label{conclusion}

Reactive solute transport in tube flows couples convection, molecular diffusion, and interphase mass transfer.
Here, the streamwise dispersion framework developed for inert solutes by \cite{guanStreamwiseDispersionSoluble2024} is extended to chemically active solutes undergoing first-order irreversible surface reaction. 
The reactive boundary destroys mass conservation and couples streamwise and radial derivatives; this derivative coupling is the central difficulty.
Unlike purely cross-sectionally averaged descriptions, the formulation retains the local spatial and streamwise transport history.
It provides analytical local and mean concentration fields and the associated transport coefficients, connecting the classical asymptotic limits with the intrinsically non-Gaussian pre-asymptotic dynamics.

The streamwise expansion is connected directly to the hierarchy of concentration cumulants. 
The exchange rate $K^S=K_0$, determined by the kinetic reactivity and the radial eigenmodes, measures the local rate of mass loss.
At early times, $K^S$ varies strongly across the tube and changes sign: the core is dominated by local depletion, whereas the near-wall region can exhibit a positive local exchange rate owing to the radial redistribution of solute.
This sign change is a transient redistribution effect.
As cross-sectional mixing proceeds, the radial contrast weakens and $K^S$ approaches a quasi-steady, radially homogenized state. 

The same redistribution determines the local drift $U^S=-K_1$ and dispersivity $D^S=K_2$. 
Because the reaction preferentially removes solute from the near-wall, low-velocity region, the surviving population is weighted towards the faster core streamlines, and the mean drift $\bar{U}_d$ increases with $k_s$. 
Removing that near-wall population reduces the contribution of slow trajectories to shear-induced spreading, so the mean dispersivity $\bar{D}_\text{eff}$ decreases with $k_s$ and approaches a reduced long-time limit below the conservative Taylor--Aris value. 
During the transitional stage, the local $D^S$ may become negative over restricted ranges of $r$ and $t$.
This does not imply a negative molecular or effective diffusivity; it reflects a transient reduction of the local second cumulant as the reactive population is redistributed among streamwise trajectories.
After cross-sectional relaxation, $D^S$ approaches a steady value, and the streamwise variance becomes asymptotically linear in time.

These coefficients account for the observed concentration patterns. 
The second-order streamwise dispersion model captures the asymmetric, skewed, and double-peaked profiles without an empirical correction, while the third- and fourth-order models modify the associated departures from Gaussianity. 
Agreement with direct Monte Carlo simulations for representative point and area sources confirms that these non-Gaussian features are captured quantitatively, so that the concentration field, its moments, and the underlying reaction--diffusion--shear mechanism are described within our streamwise framework.

The evolution separates into three regimes.  
In regime \RomanNum{1}, $t\ll t_c^{o}\sim\Pen^{-2}$, molecular diffusion dominates axial spreading and $\bar{\sigma}^2\propto 2t$, with $\alpha\simeq1$. 
In regime \RomanNum{2}, $t_c^{o}\ll t\ll t_c^{d}$, the dynamics are pre-asymptotic and controlled by shear, radial mixing, and wall reaction.
For weak reaction, $k_s\ll 1$, the dominant correction is shear-induced, and the shear-dominated transient begins at $t_c^b\sim\Pen^{-2/3}$.
For strong reaction, $k_s\gg1$, the transient is governed by the radial eigenvalue spectrum and the axial variance may grow either superlinearly or sublinearly.
The lowest radial eigenmode, with $t_c^m=\lambda_0^{-1}$, characterizes overall mass loss, whereas the next radial eigenmode, with $t_c^r(k_s)=\lambda_1^{-1}$, controls the radial relaxation that establishes the Taylor regime.
Increasing $k_s$ therefore accelerates cross-sectional relaxation, but the crossover remains finite as $k_s\to\infty$ because the ultimate rate is controlled by radial diffusion rather than by the surface reaction rate alone.
In regime \RomanNum{3}, $t\gg t_c^d\sim1$, cross-sectional equilibration is complete, the variance grows linearly and $\alpha\to1$, while $D^S\sim \mathit{O}(\Pen^2)$ with a prefactor dependent on the Damköhler number.

The small- and large-Damköhler asymptotics make this dependence explicit. 
For $k_s\ll1$, the exchange rate is reaction-limited, with $\bar K_0\to0$; for $k_s\gg1$ it approaches the diffusion-limited value $\bar K_0\to-\lim\lim{k_s\to\infty}\lambda_0\approx-5.783$.
The drift increases with $k_s$ because near-wall depletion biases the surviving solute towards the centreline and the faster streamlines.
At long times, the dispersivity decreases towards a reduced Dirichlet limit below the conservative Taylor--Aris value.
Taylor's theory \citep{taylorDispersionSolubleMatter1953} and the unsteady dispersion analyses of Gill and Sankarasubramanian \citep{gillNoteSolutionTransient1967,gillExactAnalysisUnsteady1970,sankarasubramanianUnsteadyConvectiveDiffusion1973} recover the asymptotic and Gaussian limits, whereas \cite{smithGAUSSIANAPPROXIMATIONCONTAMINANT1982}, \cite{houseworthShearDispersionResidence1984} and \cite{latiniTransientAnomalousDiffusion2001} show the difficulty of describing the transition by one Gaussian term or by separate treatments of the head and tail.
The present formulation retains the streamwise transport history and captures that non-Gaussian evolution directly. 
It is consistent with Barton's long-time global moment theory \citep{bartonAsymptoticTheoryDispersion1984}, supplies the corresponding small- and large-Damköhler limits, reduces to the conservative problem at $k_s=0$, recovers the Taylor--Aris scaling after cross-sectional equilibration, and extends continuously from that case to the Dirichlet limit.

The analysis is confined to dilute, non-interacting solutes in axisymmetric tube Poiseuille flow with a first-order irreversible surface reaction.
Within these restrictions, it describes the local exchange, drift, dispersivity, and non-Gaussian concentration evolution across all the regimes. 
Reversible reaction introduces adsorption and desorption factors \citep{boddingtonDispersionReactiveSpecies1983,jiangAnalyticalSolutionsReactive2022}. 
In porous-media channels and other confined systems with interfacial mass transfer, the cross-sectional transport operator and its eigenmodes must be adapted to the local geometry and interfacial conditions \citep{lingDispersionControlledPermeable2016,lingDispersionControlCoupled2024}.
Multi-component transport couples the species descriptions \citep{hattonDispersionMassTransfer1984,deganceTheoryDispersionChemically1985}, and curvilinear ducts, body forces, or simultaneous surface and volumetric reactions introduce further local-space couplings \citep{shapiroChemicallyReactiveGeneralized1987}.
These extensions fit the present streamwise framework, but each requires its own treatment of cross-sectional relaxation, exchange, and forcing.

\backsection[Funding]{This work is supported by the National Natural Science Foundation of China (grant nos 123B2039, 12602623, 12325207, 42450111, and 12132018), the Fellowship Programs of CPSF (grant nos GZC20251266 and 2026M791995), the International Partnership Program of the Chinese Academy of Sciences (grant no 025GJHZ2024047GC), and Young Scientists Fund Program of the Key Laboratory for Mechanics in Fluid Solid Coupling Systems, Chinese Academy of Sciences (grant no E0XM040401).}
\backsection[Declaration of interests]{The authors report no conflict of interest.}

\appendix

\section{Analytical solutions of moments}\label{appA}

A few explicit solutions for the local moments governed by (\ref{eq_moment_hierarchy}) are given below.
Further details of the analytical solution procedure can be found in \cite{bartonAsymptoticTheoryDispersion1984,biswasTaylorDispersionAbsorbing2007,wangBasicCharacteristicsTaylor2017,zhangTransientSoluteTransport2017,jiangAnalyticalSolutionsReactive2022}.

The solution for the zeroth-order local moment is
\begin{equation}
	C_0 (r,t) = 
	\sum\limits_{i=0}^{\infty} \frac{2 \lambda_i \int_0^1 Y(r) \mathrm{J}_0 (\sqrt{\lambda_i} r) r\mathrm{d}r}{(k_s^2+\lambda_i)\mathrm{J}_0^2(\sqrt{\lambda_i})} 
	\mathrm{J}_0(\sqrt{\lambda_i} r)
	\exp({-\lambda_i t}).
\end{equation}

For a central point source release, i.e. $r_0\to0$, the global zeroth moment follows by cross-sectional integration:
\begin{equation}
	M_0(t)
	=2 k_s \sum\limits_{i=0}^{\infty} 
	\dfrac{\exp({-\lambda_i t})}{(k_s^2+\lambda_i)\mathrm{J}_0(\sqrt{\lambda_i})}.
	\label{sol_M0_ring}
\end{equation}
For $t>0$, the series converges rapidly. In the limit $k_s\to0$, corresponding to an impermeable wall, $M_0=1$ and mass is conserved. For $k_s>0$, $M_0\to0$ as $t\to\infty$.

For a uniform area source, the corresponding solution is
\begin{equation}
	M_0(t) 
	=4 k_s^2 \sum\limits_{i=0}^{\infty} 
	\dfrac{\exp({-\lambda_i t})}{\lambda_i(k_s^2+\lambda_i)}.
	\label{sol_M0_area}
\end{equation}

Omitting the intermediate algebra, the first-order local moment can be written as
\begin{equation}
	C_1(r,t) = \sum\limits_{i=0}^{\infty} \Pen \overline{Y f_i} ~\overline{u f_i f_i}
	~f_i \exp({-\lambda_i t}) t
	+ \sum\limits_{i=0}^{\infty} \sum\limits_{\substack{j=0\\ j\neq i}}^{\infty}
	\Pen \dfrac{
		\overline{Y f_j}~\overline{u f_i f_j} 
		\big[\exp{(-\lambda_j t)}-\exp{(-\lambda_i t)}\big]
	}
	{\lambda_i-\lambda_j}f_i,
\end{equation}
which is equivalent to (2.9--2.13) of \cite{bartonAsymptoticTheoryDispersion1984}.

The second-order local moment is 
\begin{eqnarray}
	C_2 
	&=& \Pen^2 \sum\limits_{i=0}^{\infty} \overline{Y f_i} 
	~(\overline{u f_i f_i})^2 \exp({-\lambda_i t}) t^2 f_i
	+ \sum\limits_{i=0}^{\infty} \sum\limits_{\substack{j=0\\ j\neq i}}^{\infty}
	\dfrac{
		\overline{Y f_j}~\overline{f_i f_j}  f_i
	\big[\exp{(-\lambda_j t)}-\exp{(-\lambda_i t)}\big]
	}
	{\lambda_i-\lambda_j}\nonumber\\
	&+&\sum\limits_{i=0}^{\infty} 2\overline{Y f_i}~\overline{f_i f_i}  f_i \exp({-\lambda_i t}) t
	+\Pen^2 \sum\limits_{i=0}^{\infty} \sum\limits_{\substack{j=0\\ j\neq i}}^{\infty}
	\dfrac{
		2\overline{Y f_i}~\overline{u f_i f_j}^2  f_i \exp{(-\lambda_i t)} t
	}
	{\lambda_j-\lambda_i}
	\nonumber\\
	&+& 
	\Pen^2 \sum\limits_{i=0}^{\infty} \sum\limits_{\substack{j=0\\ j\neq i}}^{\infty} \sum\limits_{\substack{k=0\\ k\neq i,\!j}}^{\infty}
	2 \overline{Y f_k}~\overline{u f_j f_k}~\overline{u f_i f_j}  f_i
	\left[
	\dfrac{
	\exp{(-\lambda_k t)}-\exp{(-\lambda_i t)}
	}
	{(\lambda_k-\lambda_i)(\lambda_k-\lambda_j)}
	+
	\dfrac{
	\exp{(-\lambda_j t)}-\exp{(-\lambda_i t)}
	}
	{(\lambda_k-\lambda_j)(\lambda_i-\lambda_j)}
	\right]	\nonumber\\
	&+&
	\Pen^2 \sum\limits_{i=0}^{\infty} \sum\limits_{\substack{j=0\\ j\neq i}}^{\infty}
	\dfrac{
		2\overline{Y f_j}~\overline{u f_j f_j}~\overline{u f_i f_j}  f_i \exp{(-\lambda_j t)} t
	}
	{\lambda_i-\lambda_j}
	+\Pen^2	\sum\limits_{i=0}^{\infty} \sum\limits_{\substack{k=0\\ k\neq i}}^{\infty}
	\dfrac{
		2\overline{Y f_k}~\overline{u f_i f_k}~\overline{u f_i f_i}  f_i \exp{(-\lambda_i t)} t
	}
	{\lambda_k-\lambda_i}
\nonumber\\
	&+& 	
	\Pen^2 \sum\limits_{i=0}^{\infty} \sum\limits_{\substack{j=0\\ j\neq i}}^{\infty}
	\dfrac{
		2\overline{Y f_j}~\overline{u f_j f_j}~\overline{u f_i f_j}  f_i \big[\exp{(-\lambda_i t)}-\exp{(-\lambda_j t)}\big] t
	}
	{(\lambda_j-\lambda_i)^2}.
\end{eqnarray}

Higher-order solutions can be obtained analogously by separation of variables \citep{bartonMethodMomentsSolute1983}, although the resulting expressions become increasingly cumbersome and are therefore omitted.
For $n\geq2$, the formal solution can be written as
\begin{equation}
	C_n = \sum\limits_{i=0}^{\infty}
	\overline{Y f_i} f_i \exp({-\lambda_i t})+
	\int_0^{t} \bigl[n~\Pen~\overline{u f_i C_{n-1}}+ n(n-1) \overline{f_i C_{n-2}}\bigr] \exp({\lambda_i t}) \mathrm{d}t ~
	f_i \exp({-\lambda_i t}).
\end{equation}
The global moments are then obtained by cross-sectional averaging.
This recursive procedure is systematic and readily implemented numerically.
Here, only moments up to fourth order are required for the streamwise dispersion model.

\section{Numerical simulations}\label{appB}
The analytical moments are validated against direct Monte Carlo simulations.
The governing convection--diffusion equation (\ref{eq_cde}) is represented by the equivalent stochastic differential equations (SDEs). In polar coordinates, the radial dynamics are
\begin{eqnarray}
	\mathrm{d} x = \Pen ~ u(r) \mathrm{d} t + \sqrt{2} \mathrm{d} W_x,\\
	\mathrm{d} r = \frac{1}{r} \mathrm{d} t + \sqrt{2} \mathrm{d} W_r,\label{sde_r}
\end{eqnarray}
where $x$ and $r$ denote the axial and radial directions, and $W_x$ and $W_r$ are independent Wiener processes representing diffusion in the axial and radial directions.
The drift term $1/r$ introduces a numerical singularity as $r\to 0$.
To avoid this non-physical blow-up, the cross-sectional random walk is simulated in Cartesian coordinates $(y,z)$:
\begin{eqnarray}
	\mathrm{d} y = \sqrt{2} \mathrm{d} W_y, \quad \mathrm{d} z = \sqrt{2} \mathrm{d} W_z,
\end{eqnarray}
where $W_y$ and $W_z$ are independent Wiener processes, and the instantaneous radius is recovered from $r= \sqrt{y^2+z^2}$.

The irreversible mass transfer at the boundary (\ref{bc_wall}) is implemented through a stochastic partial-reaction mechanism. 
At each forward-Euler time step, a particle whose updated position lies outside the tube, $r_{\text{new}}>1$, undergoes a boundary-reaction test. 
Following stochastic particle-based implementations of continuum Robin boundary conditions \citep{andrewsAccurateParticlebasedSimulation2009, jiangAnalyticalSolutionsReactive2022,wangTransientDispersionChannel2022}, the particle is absorbed with probability
\begin{eqnarray}
	P_t = k_s \sqrt{\upi \Delta t},
\end{eqnarray}
where $\Delta t$ is the time step.
Otherwise, with probability $1-P_t$, the particle survives and is reflected back into the tube by setting $r_\text{new} \leftarrow 2-r_\text{new}$.

In the numerical simulations, we use $\Delta t=10^{-5}$ and track at least $5\times10^6$ independent particles to obtain statistically converged macroscopic moments.
Four initial spatial configurations are considered. 
In all cases, the axial position is initially localised at $x=0$, while the cross-sectional distributions are prescribed in Cartesian coordinates $(y,z)$ using a total of $N$ particles:
\begin{enumerate}
	\item \ Central point source  ($r_0\to0$): all particles are released at the centre,
	\begin{equation}
		y_\text{ini} = r_0 \cos\theta_i, \quad z_\text{ini} = r_0 \sin\theta_i.
	\end{equation}
	\item \ Uniform area source: all particles are distributed uniformly over the tube cross-section. Inverse-transform sampling is used for the radial coordinate to account for the area element,
	\begin{equation}
		y_\text{ini} = \sqrt{U_{1}} \cos(2\pi U_{2}), \quad z_\text{ini} = \sqrt{U_{1}} \sin(2\pi U_{2}).
	\end{equation}
	\item \ Internal ring source ($r_0=0.5$): all particles are placed on the concentric ring of radius:
	\begin{equation}
		y_\text{ini} = \frac{1}{2} \cos(2\pi U), \quad z_\text{ini} = \frac{1}{2} \sin(2\pi U),
	\end{equation}
	where $U\sim\mathcal{U}(0,1)$.
	\item \ Boundary ring source ($r_0=1$): all particles are released directly at the reactive boundary, maximizing their immediate interaction with the wall:
	\begin{equation}
		y_\text{ini} = \cos(2\pi U), \quad z_\text{ini} = \sin(2\pi U).
	\end{equation}
\end{enumerate}

In a reactive Monte Carlo simulation, the particle population decays over time due to interphase mass transfer. Let $N_t$ denote the number of surviving particles at time $t$, from an initial population $N$.
The global moment of order $n$ is defined by
\begin{equation}
	M_n(t) = \int_{-\infty}^{\infty} x^n \bar{C} \mathrm{d}x.
\end{equation}
In the particle-tracking framework, these moments correspond directly to ensemble averages over the surviving particles.

\begin{figure}
	\centerline{\includegraphics{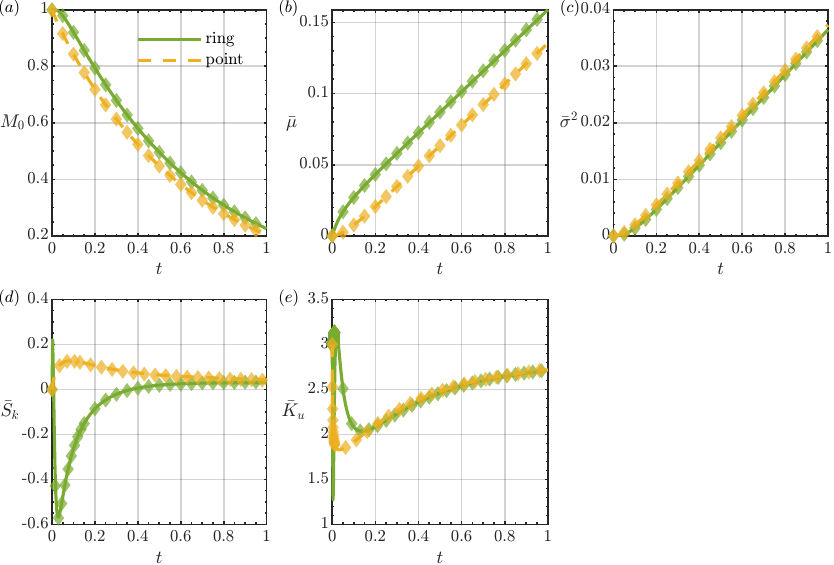}}
	\caption{Validation of the analytical solutions (lines) against direct Monte Carlo simulations (diamonds) for the global moments: ($a$) total mass, ($b$) mean, ($c$) variance, ($d$) skewness, and ($e$) kurtosis. Retaining 12 eigenfunctions is sufficient to ensure numerical convergence. The solute is initially released either from a ring source at $r=0.5$ (solid lines) or from a point source at the centre (dashed lines). The parameters are $k_s=1$, $\Pen=100$, $N=5\times10^{6}$, and $\Delta t=10^{-5}$.}
	\label{fig_verify}
\end{figure}

The zeroth moment $M_0(t)$ represents the fraction of total mass remaining in the system, and is given by
\begin{equation}
	M_0(t) = \frac{N_t}{N}.
\end{equation}
For $n \ge 1$, the raw moments $M_n(t)$ must account for the mass already removed by boundary reaction:
\begin{equation}
	M_n(t) = M_0 \cdot \left[ \frac{1}{N_t} \sum_{i=1}^{N_t} x_i^n \right] = \frac{1}{N} \sum_{i=1}^{N_t} x_i^n,
\end{equation}
where $x_i$ is the axial position of the $i$-th particle.

The first two normalised cumulants are the mean and variance of the surviving solute cloud. 
The mean centroid is
\begin{equation}
	\bar\mu(t) = \frac{M_1}{M_0} = \frac{1}{N_t} \sum_{i=1}^{N_t} x_i.
\end{equation}
The mean variance is
\begin{equation}
	\bar\sigma^2(t) = \frac{M_2}{M_0} - \left[ \frac{M_1}{M_0} \right]^2 = \frac{1}{N_t} \sum_{i=1}^{N_t} \left( x_i - \bar\mu \right)^2.
\end{equation}

In the presence of interphase mass transfer ($k_s > 0$), the effective mean dispersivity calculated from these particle cumulants through the Taylor--Aris relation,
\begin{equation}
	\bar D_{\text{eff}} = \frac{1}{2} \frac{\mathrm{d}\bar\sigma^2}{\mathrm{d}t},
\end{equation}
approaches an asymptotic constant determined by the coupled effects of shear convection, radial diffusion and boundary-induced depletion.

The skewness characterizes the asymmetry of the surviving particle distribution about its moving centroid. It is defined by
\begin{equation}
	\bar S_k(t) = \dfrac{{M_3}/{M_0} - 3\bar\mu M_2/M_0 +2 \bar\mu^3}{\bar\sigma^3} = \dfrac{\frac{1}{N_t} \sum\limits_{i=1}^{N_t} 
	\left( x_i - \bar\mu \right)^3}{\left[ \frac{1}{N_t} \sum\limits_{i=1}^{N_t} \left( x_i - \bar\mu \right)^2 \right]^{3/2}}.
\end{equation}

The kurtosis is the fourth central moment normalized by the squared variance:
\begin{equation}
	\bar K_u(t) = \frac{{M_4}/{M_0} - 4\bar\mu{M_3}/{M_0} - 3 M_2/M_0 + 12\bar\mu^2 {M_2}/{M_0} - 6\bar\mu^4}{\bar\sigma^4}= \frac{\frac{1}{N_t} \sum\limits_{i=1}^{N_t} \left( x_i - \bar\mu \right)^4}{\left[ \frac{1}{N_t} \sum\limits_{i=1}^{N_t} \left( x_i - \bar\mu \right)^2 \right]^2}.
\end{equation}

The analytical global moments derived in Appendix~\ref{appA} are validated against direct Monte Carlo simulations in figure~\ref{fig_verify}. 
The total mass is obtained explicitly for the different initial conditions from (\ref{sol_M0_ring})--(\ref{sol_M0_area}), as shown in figure~\ref{fig_verify}($a$). 
Good agreement is obtained for all five global moments shown.

\bibliographystyle{jfm}
\bibliography{mylib}

\end{document}